\documentclass[preprintnumbers,superscriptaddress,endnote,nofootinbib,aps,prd,floatfix,11pt]{revtex4}

\usepackage[utf8]{inputenc}

\usepackage{footmisc,enumerate}
\usepackage{multirow}
\usepackage{subfigure}
\usepackage{amsmath,amssymb}
\usepackage{graphicx}
\usepackage{placeins}
\usepackage{bbm}
\usepackage{xspace,slashed}
\usepackage{hyperref}
\hypersetup{colorlinks=true, citecolor=blue, urlcolor=blue, linkcolor=blue}
\usepackage[normalem]{ulem}

\usepackage{amsmath}
\usepackage{amsfonts}
\usepackage{amssymb}
\usepackage{graphicx}
\usepackage{cancel}
\usepackage{xcolor}
\usepackage{hyperref}
\usepackage{multirow}
\usepackage{pifont}

\usepackage{threeparttable}

\usepackage[autostyle]{csquotes}

\newcommand{\dphiX}{\ensuremath{\Delta\phi_{X}}\xspace}

\newcommand{\tx}[1]{{#1}}

\newcommand{\cmark}{\ding{51}}%
\newcommand{\xmark}{\ding{55}}%
\newcommand{\D}{D\mkern-11.5mu/}

\begin{document}

\newcommand{\og}{\ensuremath{\tilde{O}_g}\xspace}
\newcommand{\ot}{\ensuremath{\tilde{O}_t}\xspace}

\providecommand{\abs}[1]{\lvert#1\rvert}

\newcommand{\Znunujets}{(Z\to{\nu\bar{\nu}})+\text{jets}}
\newcommand{\Welnujets}{(W\to{\ell\nu})+\text{jets}}
\newcommand{\Znunujet}{(Z\to{\nu\bar{\nu}})+\text{jet}}
\newcommand{\Welnujet}{(W\to{\ell\nu})+\text{jet}} 

\newcommand{\cw}{\ensuremath{C_{\widetilde{W}}\xspace}}
\newcommand{\chwb}{\ensuremath{C_{H\widetilde{W}B}}\xspace}


\newcommand{\ce}[1]{{\color{red}{#1}}}
\newcommand{\jd}[1]{{\color{red}{#1}}}


\title{ATLAS Violating CP Effectively}

\begin{abstract}
CP violation beyond the Standard Model (SM) is a crucial missing piece for explaining the observed matter-antimatter asymmetry in the Universe. Recently, the ATLAS experiment at the Large Hadron Collider has performed an analysis of electroweak $Zjj$ production, thereby excluding the SM locally at 95\% confidence level in the measurement of CP-sensitive observables. We take
the excess' interpretation in terms of anomalous gauge-Higgs interactions at face value and discuss further steps that are required to scrutinize its origin. In particular,
we discuss the relevance of multi-boson production using adapted angular observables to show how they can be used to directly tension the reported $Zjj$ excess in a more comprehensive analysis. To connect the excess to a concrete UV scenario for which the underlying assumptions of the $Zjj$ analysis are valid, we identify vector-like leptons as a candidate theory consistent with the observed CP-odd Wilson coefficient hierarchy observed by ATLAS. We perform a complete one-loop matching calculation to motivate
further model-specific and correlated new physics searches. In parallel, we provide estimates of the sensitivity reach of the LHC's high luminosity phase for this particular scenario of CP-violation in
light of electroweak precision and Run-2 Higgs data. These provide strong constraints on the model's CP-even low-energy phenomenology, but also inform the size of the CP-odd SM deformation indirectly via our model hypothesis. 
\end{abstract}


\author{Supratim~Das~Bakshi} \email{sdbakshi@iitk.ac.in}
\affiliation{Indian Institute of Technology Kanpur, Kalyanpur, Kanpur 208016, India\\[0.1cm]}
\author{Joydeep~Chakrabortty} \email{joydeep@iitk.ac.in}
\affiliation{Indian Institute of Technology Kanpur, Kalyanpur, Kanpur 208016, India\\[0.1cm]}
\author{Christoph~Englert} \email{christoph.englert@glasgow.ac.uk}
\affiliation{School of Physics \& Astronomy, University of Glasgow, Glasgow G12 8QQ, United Kingdom\\[0.1cm]}
\author{Michael~Spannowsky} \email{michael.spannowsky@durham.ac.uk}
\affiliation{Institute for Particle Physics Phenomenology, Durham University, Durham DH1 3LE, United Kingdom\\[0.1cm]}
\author{Panagiotis~Stylianou}\email{p.stylianou.1@research.gla.ac.uk} 
\affiliation{School of Physics \& Astronomy, University of Glasgow, Glasgow G12 8QQ, United Kingdom\\[0.1cm]}

\preprint{IPPP/20/43}

\maketitle

\section{Introduction}
\label{sec:intro}
The search for non-Standard Model (SM) sources of CP violation is a crucial missing
piece in connecting the phenomenological success of the SM so far with its apparent
shortcomings related to the observed baryon anti-baryon asymmetry~\cite{Sakharov:1967dj}. 
Searches for CP violation in various channels at the Large Hadron Collider (LHC) are therefore
a key part of the ongoing experimental program (see e.g. \cite{Aad:2020ivc,Sirunyan:2020sum} for recent analyses in the context of Higgs physics).

In particular, ATLAS has recently performed a detailed analysis of electroweak $Z+2j$ production in Ref.~\cite{Aad:2020sle}, where
it also interprets measurements in terms of effective field theory (EFT) deformations of the SM using dimension six CP-violating operators
in the Warsaw basis,~\cite{Grzadkowski:2010es}
\begin{align}
\label{eq:ops}
Q_{\widetilde W} & = \epsilon^{abc}\widetilde{W}^{a}_{\mu\nu }W^{b\,\nu\rho} W^{c\,\rho\mu}\,,\\
Q_{H\widetilde W B} & = (H^\dagger\tau^a H)\widetilde{W}_{\mu\nu}^a B^{\mu\nu}\,,
\end{align}
where $W,B$ denote the field strengths of weak $SU(2)_L$ and hypercharge $U(1)_Y$, $H$ is the Higgs doublet, $\tau^a$ are the Pauli matrices, and the tilde refers to the dual field strength tensor $\widetilde{X}_{\mu\nu}= \epsilon_{\mu\nu\delta\rho} X^{\delta\rho}/2$ ($X=W,B,G$). Using the effective Lagrangian
\begin{equation}
\label{eq:atlaslag}
{\cal{L}}= {\cal{L}}_{\text{SM}} + {C_{\widetilde W}\over \Lambda^2}  Q_{\widetilde W}  + {C_{H\widetilde W B} \over \Lambda^2} Q_{H\widetilde W B} \, ,
\end{equation}
ATLAS provides the observed 95\% confidence level constraints on the following CP violating operators~\cite{Aad:2020sle}
\begin{equation}
\label{eq:tension}
{C_{\widetilde W}} {\text{TeV}^2\over \Lambda^2} \in [-0.11, 0.14] \,,\quad  {C_{H\widetilde W B}}  {\text{TeV}^2\over \Lambda^2} \in [0.23, 2.34]\,,
\end{equation}
based on dimension six interference-only contributions arising from matrix elements
\begin{equation}
\label{eq:matrix}
|{\cal{M}}|^2 = |{\cal{M}}_{\text{SM}}|^2 + 2\, \text{Re} \left[ {\cal{M}}_{\text{SM}}\,  {\cal{M}}_{\text{d6}}^\ast (  C_{\widetilde W},C_{H\widetilde W B})\right].
\end{equation}
This leads to asymmetries in $P$-sensitive distributions, such as the ``signed'' (according to rapidity) azimuthal angle difference of the tagged jets $\Delta\Phi_{jj}$. The benefit of such observables and the study of their asymmetries is that the CP-even deformations do not contribute to the exclusion constraints directly, which also extends to CP-even modifications arising from ``squared'' dimension six contributions. In Eq.~\eqref{eq:matrix}, ${\cal{M}}_{d6}$ denotes the amplitude contribution  from the operators of Eq.~\eqref{eq:ops}, thus it is a linear function of $ C_{\widetilde W}/\Lambda^2,C_{H\widetilde W B}/\Lambda^2$ (as we are keeping terms up to order $1/\Lambda^2$).

The constraint on the Wilson coefficient $C_{H\widetilde W B}$ in Eq.~\eqref{eq:tension} indicates a tension with the SM while the observed cross section agrees well with the SM expectation with $39.5$~fb~data~\cite{Arnold:2008rz,Baglio:2014uba,Bellm:2015jjp}. This prompts us to the following interesting questions. 

Firstly, the tension of Eq.~\eqref{eq:tension} seems to rule out the SM at an SM-compatible cross sections. Experimental analyses of asymmetries are challenging, and systematics are crucial limiting factors of distribution shape analyses. Nonetheless, the result of Ref.~\cite{Aad:2020sle} could indeed be the first glimpse of a phenomenologically required and motivated extension of the SM, thus deserving further experimental and theoretical scrutiny. 

Secondly, limiting ourselves to a subset of the dimension six operators that could in principle contribute to physical process can be theoretically problematic, in particular when we wish to interpret the experimental findings in a truly model-independent fashion. While concrete UV scenarios can be expected to exhibit hierarchical Wilson coefficient patterns, it is not a priori clear that limiting oneself to anomalous gauge boson interactions has a broad applicability to UV scenarios. 

Addressing these two questions from a theoretical and phenomenological perspective is the purpose of this work. In Sec.~\ref{sec:scrut}, we motivate additional diboson analyses of the current ${\cal{O}}(100)~\text{fb}^{-1}$ data set that will allow us to tension or support the results of Eq.~\eqref{eq:tension} straightforwardly. This is particularly relevant as the ATLAS constraints amount to a large, and as it turns out non-perturbative, amount of CP violation associated with a single direction in the EFT parameter space. In Sec.~\ref{sec:theory}, we show that the ATLAS assumptions of considering two operators are consistent for models of vector-like leptons, which can not only reproduces a hierarchy $|C_{H\widetilde W B}|/\Lambda^2 > |C_{\widetilde W}|/\Lambda^2$ as suggested by Ref.~\cite{Aad:2020sle}, but also collapse the analysis-relevant operators to those modifying the gauge boson self-interactions for the considered analyses. Combining both aspects, in Sec.~\ref{sec:extrapo} we assess the future of diboson and $Z+2j$ analyses from a perturbative perspective and discuss the high-luminosity (HL) sensitivity potential of the LHC in light of the electroweak precision constraints. We conclude in Sec.~\ref{sec:conc}.

\section{Scrutinizing $C_{\widetilde{HWB}}$ with diboson production and current LHC data}
\label{sec:scrut}
Deviations related to the gauge boson self-coupling structure can be scrutinized using abundant diboson production at the LHC. With clear leptonic final states and large production cross sections, these signatures are prime candidates for electroweak precision analyses in the LHC environment with only a minimum of background pollution, see also~\cite{Aad:2011tc,Chatrchyan:2013fya}. In particular, radiation zeros observed in $W\gamma$ production are extremely sensitive to perturbations of the SM CP-even coupling structures~\cite{Goebel:1980es,Brodsky:1982sh,Brown:1982xx,Baur:1993ir,Baur:1994sa,Han:1995ef,Aihara:1995iq}. In this section, we discuss the relevant processes that can be employed to further tension the findings of Eq.~\eqref{eq:tension}.

\subsection{Processes}
\label{sec:process}
The squared amplitude of Eq.~\eqref{eq:matrix}
receives interference contributions from dimension six operators that in the special case where they are CP-odd, do not change the cross section of a process but appear in CP-sensitive observables.  Anomalous weak boson interactions were studied in Ref.~\cite{Aad:2020sle} through the $Zjj$ channel by the introduction of two CP-violating operators, $Q_{\widetilde{W}}$ and $Q_{H \widetilde{W} B}$, modifying the differential distribution of the parity-sensitive signed azimuthal angle between the two final state jets $\Delta \phi_{jj} = \phi_{j_1} - \phi_{j_2}$, where $\phi_{j_1}$ ($\phi_{j_2}$) is the azimuthal angle of the first (second) jet, as ordered by rapidity. Similar parity-sensitive observables can be constructed for the leptonic final states of the $W \gamma \to \ell \nu \gamma$, $W^+ W^- \to \ell^+ \nu_\ell \ell^- \bar{\nu}_\ell$, and $W Z \to \ell \nu \ell^+ \ell^-$ channels allowing to further constrain the reach of the two Wilson coefficients. 

The operators are modeled using {\sc{FeynRules}}~\cite{Christensen:2008py,Alloul:2013bka} and exporting the interactions through a {\sc{UFO}}~\cite{Degrande:2011ua} file. Events are generated using {\sc{MadEvent}}~\cite{Alwall:2011uj,deAquino:2011ub,Alwall:2014hca} through the {\sc{MadGraph}} framework~\cite{Alwall:2014hca} and saved in the {\sc{LHEF}} format~\cite{Alwall:2006yp}, before imposing selection criteria and cuts. 

\subsection*{$WZ$ production at the LHC}
We study the $WZ$ channel by selecting leptons in the pseudorapidity $|\eta(\ell)| < 2.5$ and transverse momentum $p_T > 5$~GeV regions. Exactly three leptons are required and at least one same-flavor opposite-charge lepton pair must have an invariant mass within the $Z$ boson mass window $m_{\ell\ell} \in \left[ 60, 120 \right]$~GeV. In the case of more than one candidate pairs, the one that yields an invariant mass closest to the $Z$ boson is selected. The remaining lepton $\ell^\prime$ is required to have $p_{T} ( \ell^\prime ) > 20$~GeV. To obtain a P-sensitive observable, we reconstruct the dilepton pair four-momentum and obtain the rapidity $y_{\ell \ell}$ and azimuthal angle $\phi_{\ell \ell}$. We order the dilepton and third lepton azimuthal angles based on the rapidities of the two reconstructed objects, such that $\phi_1$ ($\phi_2$) is the one with the greatest (smallest) rapidity. The signed azimuthal angle is then constructed as $\Delta \phi_{\ell^\prime Z} = \phi_1 - \phi_2$. 

The distributions of the signed azimuthal angle for both the SM and the SM-BSM interference are normalized to the CMS measured fiducial cross section~\cite{Khachatryan:2016tgp} of the particular phase space region at $13$~TeV center of mass energy  
\begin{equation}
	\label{eq:wz_xsec}
	\sigma_{\text{fid}} (p p \to W Z \to \ell^\prime \nu \ell \ell) = 258 \pm 21 (\text{stat}) ^{+19}_{-20} (\text{syst}) \pm 8.0 (\text{lumi}) \;\text{fb}\;. 
\end{equation}

\subsection*{$WW$ production at the LHC}
Turning to the $WW$ channel and following Ref.~\cite{Aaboud:2019nkz}, we produce events decaying to the $WW \to e \nu_e \mu \nu_\mu$ final state. The two leptons $e$ and $\mu$ are required to satisfy $|\eta(\ell)| < 2.5$ and $p_T(\ell) > 27$~GeV with no third lepton in the $p_T > 10$~GeV region. Contributions from the Drell-Yan background are reduced by imposing cuts on the missing energy $E_T > 20$~GeV and on the transverse momentum of the dilepton pair $p_T(e \mu) > 30$~GeV. The phase space region is constrained further by enforcing the invariant mass condition $m(e \mu) > 55$~GeV that suppresses the $H \to WW$ background. In this channel the signed azimuthal angle $\Delta \phi_{\ell\ell}$ is then defined directly from the azimuthal angles of the two leptons sorted by rapidity. 

The fiducial cross section of $WW \to e \mu + {\slashed{E}}_T$ was measured by ATLAS~\cite{Aaboud:2019nkz} as 
  \begin{equation}
	  \label{eq:ww_xsec}
	  \sigma_{\text{fid}} (p p \to W W \to \ell \nu_e \mu \nu_\mu) = 379.1 \pm 5.0 (\text{stat}) \pm 25.4 (\text{syst}) \pm 8.0 (\text{lumi}) \; \text{fb}\; , 
  \end{equation}
which is used to normalize the calculated differential distribution of $\Delta \phi_{\ell\ell}$. The total cross section of the events as well as the relative statistical and systematic uncertainties are subsequently rescaled to include the final states of all light leptons $WW \to \ell \nu \ell \nu$. 

\subsection*{$W \gamma$ production at the LHC}
To obtain the cross section of $W \gamma$ at $13$~TeV, we first use {\sc{MCFM}}~\cite{Campbell:1999ah,Campbell:2011bn,Campbell:2015qma,Boughezal:2016wmq,Campbell:2019dru} with generation level cuts $p_T > 10$~GeV and $|\eta| < 2.5$ for both leptons and photons, requiring the separation $\Delta R (\ell, \gamma) > 0.4$, in order to obtain the cross section at NLO precision with $p_T(\gamma)$ as the renormalization and factorization scale. We have validated these choices against early measurements from ATLAS~\cite{Aad:2011tc} and CMS~\cite{Chatrchyan:2013fya}. The events are generated as before with {\sc{MadEvent}} using the same generation cuts and we rescale the computed {\sc{MadEvent}} cross section of the events to the {\sc{MCFM}} value, in order to include higher order effects and obtain normalized distributions. 

Post-generation we veto events without at least one lepton (photon) with transverse momentum $p_T(\ell) > 35$~GeV ($p_T(\gamma) > 15$~GeV and require a separation of $\Delta R (\ell, \gamma) > 0.7$. The azimuthal angles of the photon and the lepton are sorted by rapidity and $\Delta \phi_{\ell \gamma}$ is calculated similarly to the other channels.

We assume that the relative statistical and systematic errors that can be calculated from the measured cross section of Ref.~\cite{Chatrchyan:2013fya} \footnote{$\ell$ for this cross section indicates each type of light lepton ($e$, $\mu$) and not a sum over them.}
\begin{equation}
	\label{eq:wa_xsec7tev}
	\sigma_{\text{fid}}( p p \to W \gamma \to \ell \nu) = 37.0 \pm 0.8 (\text{stat}) \pm 4.0 (\text{syst}) \pm 0.8 (\text{lumi}) \; \text{pb}, 
\end{equation}
will remain the same for the case of $\sqrt{s} = 13$~TeV and use this in the following statistical analysis.


\subsection{Analysis of CP-sensitive observables}
To study the allowed region of the $(\cw, \chwb)$ parameter space based on current experimental data at the LHC, we consider the differential distribution
\begin{equation}
	\label{eq:diffdist}
	\frac{\tx{d}\sigma(\cw,\chwb)}{\tx{d}\dphiX}=\frac{\tx{d}\sigma_{\tx{SM}}}{\tx{d}\dphiX}+\cw\frac{\tx{d}\sigma_{\widetilde{W}}}{\tx{d}\dphiX}+\chwb \frac{\tx{d}\sigma_{H\widetilde{W}B}}{\tx{d}\dphiX}\,,
\end{equation}
where, depending on the process, $X = \ell^\prime Z, \ell\ell, \ell\gamma$, and $\sigma_{H\widetilde WB}$ and $\sigma_{\widetilde W}$ are constructed from $Q_{H \widetilde{W} B}$ and $Q_{\widetilde{W}}$, respectively, and derive from MC integration of Eq.~\eqref{eq:matrix}. We generate events for each process using the two coupling reference points $(\cw, \chwb) = (1,0)$ and $(\cw, \chwb) = (0,1)$ and can rescale distributions using the linear relation of~Eq.~\eqref{eq:matrix} to subsequently scan over the space of the two CP-odd Wilson coefficients, performing a $\chi^2$ fit, in order to obtain limits. The $\chi^2$ statistics is defined as
\begin{equation}
	\label{eq:chi2}
	\chi^2(\cw,\chwb) =\big(b_{\tx{SM+d6}}^i(\cw,\chwb) - b_{\tx{SM}}^i\big)V_{ij}^{-1}\big( b_{\tx{SM+d6}}^j(\cw,\chwb) -  b_{\tx{SM}}^j \big)\,,
\end{equation}
\begin{figure}[!b]
\begin{center}
   \parbox{0.55\textwidth}{\includegraphics[width=0.55\textwidth]{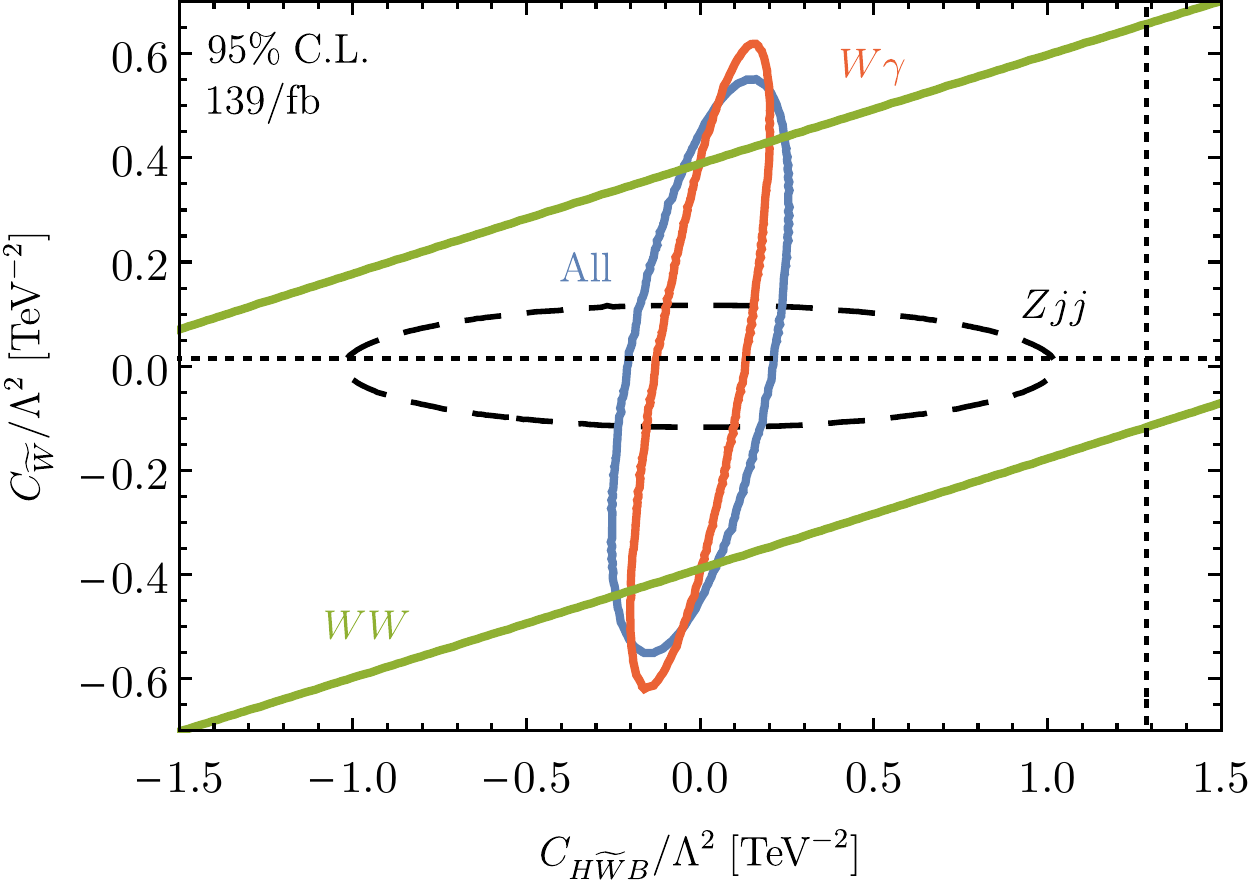}}
   \hfill
   \parbox{0.39\textwidth}{
   \vspace{-0.cm}
	\caption{Exclusion contours for $W\gamma$  and $WW$ are shown separately and when combined for $139/$fb. $WZ$ does not provide significant sensitivity and lies outside the plotting region. We overlay the diboson constraints with the $Z+2j$ as extracted from the confidence intervals of ATLAS and the best fit lines (dotted) from experimental observations~\cite{Aad:2020sle}. \label{fig:excl139}}}
\end{center}
\end{figure}
where $b_{\tx{SM+d6}}^i(\cw,\chwb)$ is the number of events at a particular luminosity based on the $i^\text{th}$ bin of the differential distribution~Eq.~\eqref{eq:diffdist} for a set of Wilson coefficients and $b_{\tx{SM}}^i$ is the bin's expected number of events based solely on the SM. The covariance matrix $V_{ij}$ includes the relative statistical and systematic uncertainties\footnote{Luminosity uncertainties are treated as systematics.} from the experimental measurements, obtained from the aforementioned fiducial cross sections~Eqs.~\eqref{eq:wz_xsec}, \eqref{eq:ww_xsec}, and \eqref{eq:wa_xsec7tev} for each process and included in $V_{ij}$ as terms of the form  $(\varepsilon_{\text{rel. stat.}}^2 + \varepsilon_{\text{rel. syst}}^2) b_{\tx{SM}}^i b_{\tx{SM}}^j$, assuming that both relative and systematic errors are fully correlated.  $\varepsilon_{\text{rel. stat.}}$ and $\varepsilon_{\text{rel. stat.}}$ denote the relative statistical and systematic uncertainties of each process. 

We define the confidence intervals with 
\begin{equation}
	\label{eq:pval}
	1 - CL \geq \int_{\chi^2}^{\infty}  {\text{d}}x \,p_k(x)\,,~\quad \chi^2=\chi^2(\cw/\Lambda^2,\chwb/\Lambda^2)\,,
\end{equation}
using the $\chi^2$ distribution of $k$ degrees of freedom $p_k(x)$, where $k$ is obtained by subtracting the number of Wilson coefficients from the number of measurements. 

We perform a scan based on an integrated luminosity of $139/$fb to obtain the $95\%$ confidence level contours shown in Fig.~\ref{fig:excl139}. The results are overlapped with the $Z+2j$ allowed region from ATLAS~\cite{Aad:2020sle}, as well as the best fit point from experimental data, while the $WZ$ does not constrain the region enough to appear on the plot. To obtain the $Z+2j$ contours, we have tuned a covariance matrix on the basis of the information of Ref.~\cite{Aad:2020sle} to obtain the exclusions reported in their work.

As can be seen the measurement of $Z+2j$ is considerably more sensitive to $Q_{\widetilde W }$ than to $Q_{H\widetilde W B}$, which results from a combination of accessing $t$-channel momentum transfers in the weak boson fusion-type selections and the $Z$ boson having a larger overlap with the $W^3$ field than the photon. The latter is also the reason why $W\gamma$ production enhances the sensitivity in the $Q_{H\widetilde W B}$ direction. We note that electroweak mono-photon production in association with two jets is more challenging due to jet-misidentification, and thus does not provide significant sensitivity compared to prompt $W\gamma$ production.

\subsection{HL-LHC extrapolation}
\begin{figure}[!t]
\begin{center}
   \parbox{.55\textwidth}{\includegraphics[width=0.55\textwidth]{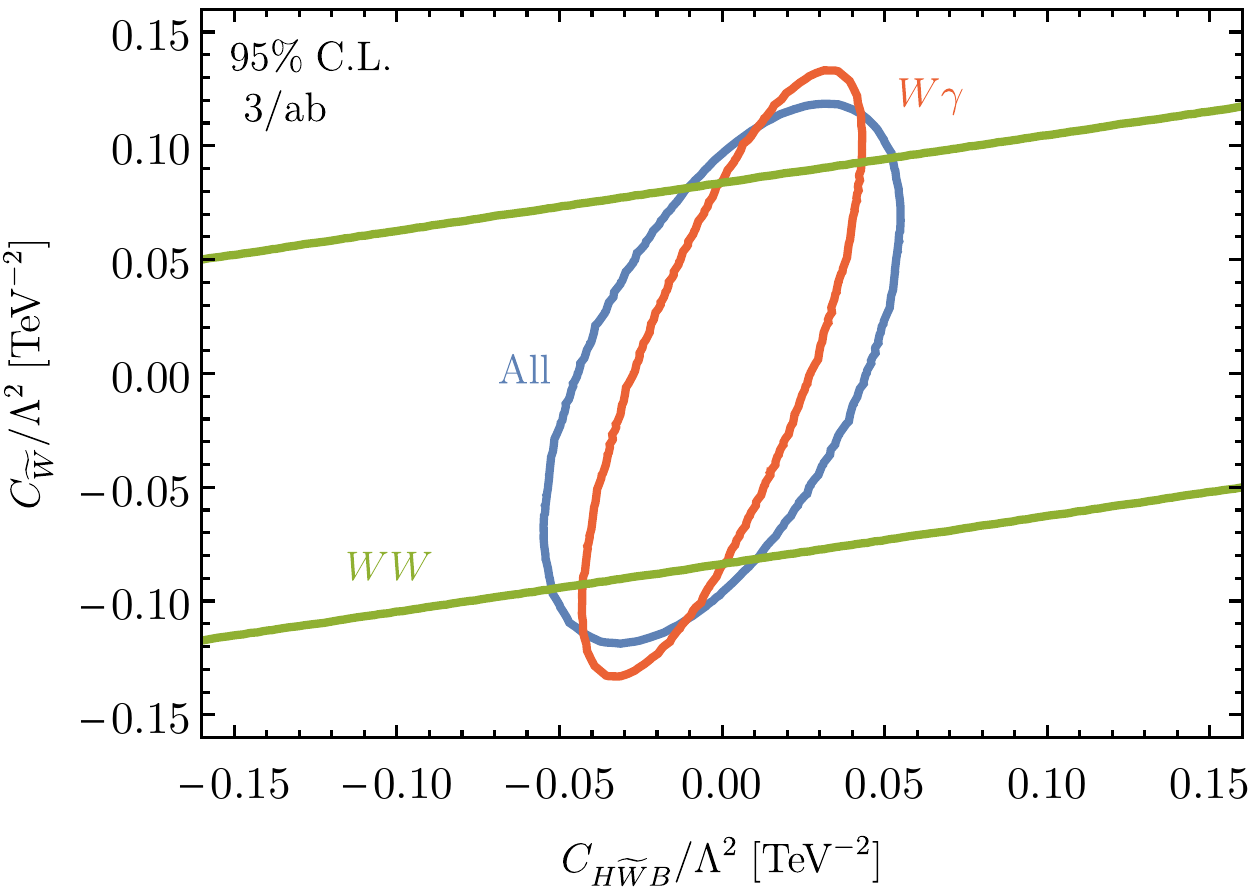}}
   \hfill
   \parbox{0.39\textwidth}{
   \vspace{-1.5cm}
	\caption{Same as Fig.~\ref{fig:excl139}, but extrapolated to an integrated HL-LHC luminosity of 3/ab. The contours depend mostly on the statistical fluctuations and no significant change occurs when statistical errors are reduced.\label{fig:excl3ab}}}
\end{center}
\end{figure}

We repeat the analysis with the same technique but using an integrated luminosity of $3/$ab to obtain contours for HL-LHC. Systematic errors could be significantly reduced at HL-LHC, however for this particular case we find that the fact that BSM contributions are antisymmetric functions, in contrast to the symmetric SM differential distribution, leads to cancellations of the introduced errors in the $\chi^2$.\footnote{The same occurs if the absolute systematic errors are distributed according to a symmetric shape distribution across the bins, instead of using relative errors.} Hence, the analysis is predominantly limited only by the statistical fluctuations. The extrapolated contours for $3/$ab are shown in Fig.~\ref{fig:excl3ab}.

\section{Vector-like leptons as a model for $C_{\widetilde{HWB}}$-targeted (di-)boson signals}
\label{sec:theory}
Let us return to the examining the excess related to $Q_{H\widetilde W B}$ from a UV model perspective. To this end, we extend the SM by three heavy vector-like lepton (VLL) multiplets \cite{Angelescu:2020yzf}
\begin{equation}
	 \Sigma_{L,R}=\begin{pmatrix} \eta \\ \xi	\end{pmatrix}_{L,R} : (1,2,\mathcal{Y}), \;\;
	 \eta'_{L,R} : (1,1,\mathcal{Y}+\frac{1}{2}),\;\;
	 \xi'_{L,R} : (1,1,\mathcal{Y}-\frac{1}{2}),
\end{equation}
where the quantum numbers are depicted in $SU(3)_C\otimes SU(2)_L \otimes U(1)_Y$ convention.

The most-general gauge-invariant renormalizable Lagrangian involving these heavy VLLs can be written as
\begin{multline}
\label{eq:vlike}
	\mathcal{L}_{\text{VLL}}  =  \bar{\Sigma} ( i \D_{_\Sigma} - m_{_{\Sigma}}) \Sigma + \bar{\eta'} ( i \D_{\eta'} - m_{{\eta'}}) \eta' + \bar{\xi'} ( i \D_{\xi'} - m_{{\xi'}}) \xi' \\
	- \left\lbrace \bar{\Sigma} \tilde{H} ( Y_{\eta_{_L}} \mathbb{P}_L + Y_{\eta_{_R}} \mathbb{P}_{R}) \eta' + \bar{\Sigma} H ( Y_{\xi_{_L}} \mathbb{P}_{L} + Y_{\xi_{_R}} \mathbb{P}_{R}) \xi' + \text{h.c.} \right\rbrace,
\end{multline}
where,
$m_{_{\Sigma}}, m_{\eta'}$, and $m_{\xi'}$ are the masses of $\Sigma$, $\eta'$ and $\xi'$, respectively. $\mathbb{P}_{L}(\mathbb{P}_{R})$ are the left (right) chiral projection operator. $Y_i$'s are the complex Yukawa couplings. We will consider $m_{_\Sigma} = m_{\eta'} = m_{\xi'} = m$, i.e., all the VLLs are degenerate  in this work.\footnote{This model is also discussed in some detail in Refs.~\cite{Corbett:2017ecn,Bakshi:2018ics,Angelescu:2020yzf}.} We will see that this class of models provides the appropriate UV backdrop of for the Wilson coefficient analysis that we have performed above, and on which Ref.~\cite{Aad:2020sle} relies.

\subsection{Wilson Coefficients}
We integrate out all three heavy degenerate VLL multiplets, see  Eq.~(\ref{eq:vlike}) leading to the effective Lagrangian
\begin{align}\label{eq:eft}
	\mathcal{L}_{\text{EFT}}= \mathcal{L}_{\text{SM}} + \frac{1}{16 \pi^2 m^2} \sum\limits_i  {\mathcal{C}}_i Q_i \, ,
\end{align}
where $Q_i$, ${\mathcal{C}}_i$ denote the effective dimension six operators and the Wilson coefficients respectively. The UV theory in Eq.~(\ref{eq:vlike}) is suitably matched to the SMEFT  at the scale $m$ which serves as the cut-off scale of the EFT. Here, the $16\pi^2$ factor signifies that all the effective operators are generated through one-loop.  And we separate off the loop factor $(16\pi^2)^{-1}$ from the definition of the Wilson coefficients ${\cal{C}}_i$, i.e. $C_i= {\cal{C}}_i/16\pi^2 $ and $\Lambda=m$ in comparison with Eq.~\eqref{eq:atlaslag}. We employ the $\overline{\text{MS}}$ renormalization scheme and also set the RG scale at $\mu=m$. Integrating out heavy fermions from UV theories is discussed in Refs.~\cite{Angelescu:2020yzf, Ellis:2020ivx}. Note that dimension eight CP-violating effects play a subdominant role when perturbative matching is possible in the first place, see~\cite{Belusca-Maito:2017iob,Corbett:2017ecn}. We can therefore expect the dimension six deformations to play a dominant role.

We present the effective operators in the Warsaw basis \cite{Grzadkowski:2010es} and their respective Wilson coefficients (WCs) are encapsulated in Tab.~\ref{tab:warsaw1}. We also provide the matching using strongly-interacting Light Higgs (SILH)-like convention~of~\cite{Angelescu:2020yzf} (see also \cite{Giudice:2007fh,Contino:2013kra}) in Tab.~\ref{SILH} of Appendix~\ref{sec:silh} for convenience. We compute  the most generic results using the complete Lagrangian including CP conserving and violating interactions simultaneously. A subset of our generic results (in SILH-like basis) is in well agreement with operators computed in Ref.~\cite{Angelescu:2020yzf}. The results in WARSAW for this VLL scenario has been computed for the first time in this paper.\footnote{In Ref.~\cite{Angelescu:2020yzf} the contributions from CP violating (CPV) couplings into the CP-even operators are not considered.} We find 19 effective operators with non-zero Wilson coefficients (16 CP-even + 3 CP-odd). In the renormalizable Lagrangian, the VLLs interact with the SM Higgs doublet and that explains the origin of 10 bosonic  along with 9 fermionic effective operators accompanied by non-zero WCs. These appear due to application of the equation of motion of the SM Higgs doublet on the effective Lagrangian. 
\begin{table*}[p]
	\caption{The generated Warsaw basis operators and respective Wilson coefficients after integrating out VLLs using Eq.~\eqref{eq:vlike}. The  CP-odd gauge boson operators are displayed in first three rows. Multiplication with a common factor $({16 \pi^2 m^2})^{-1}$ is understood implicitly, see Eq.~(\ref{eq:eft}).}
	\label{tab:warsaw1}
	\vskip 0.2cm
	\footnotesize
	\centering
	\renewcommand{\arraystretch}{1.2}
	\begin{threeparttable}
	\begin{tabular}{|c|c|l|}
		\hline\hline
		$\text{Operators}$&$\text{Operator Structures}$&$\text{Wilson Coefficients} \ \left({\cal{C}}_i\right)$\\
		\hline\hline
		$Q_{H\widetilde{B}}$  &  $\left(H^{\dagger }H^{  }\right)\widetilde{B}_{\mu \nu }B^{\mu \nu }$  &   $-\frac{g_{_Y}^2}{12}\left[(1+6\mathcal{Y}+12\mathcal{Y}^2)\text{Im}[Y_{\eta_{_L}} Y^*_{\eta_{_R}}] +(1-6\mathcal{Y}+12\mathcal{Y}^2) \text{Im}[Y_{\xi_{_L}} Y^*_{\xi_{_R}}]\right]$  \\
			\hline
		$Q_{H\widetilde{W}}$  &  $\left(H^{\dagger }H^{  }\right)\widetilde{W}_{\mu \nu }{}^aW^{a,\mu \nu }$  &  $-\frac{g_{_W}^2}{12}\text{Im}[Y_{\eta_{_L}}Y^*_{\eta_{_R}}+Y_{\xi_{_L}}Y^*_{\xi_{_R}}]$  \\
		\hline
		$Q_{H\widetilde{W}B}$  &  $\left(H^{\dagger }\tau ^aH^{  }\right)\widetilde{W}_{\mu \nu }{}^aB^{\mu \nu }$  & $\frac{g_{_W} g_{_Y}}{6}\left[ (1+6\mathcal{Y}) \text{Im}[Y_{\eta_{_L}} Y^*_{\eta_{_R}}] + (1-6\mathcal{Y}) \text{Im}[Y_{\xi_{_L}} Y^*_{\xi_{_R}}] \right]$ \\
		\hline
		\hline
		$Q_W$  &  $\epsilon ^{{abc}}W_{\rho }{}^{a,\mu }W_{\mu }{}^{b,\nu }W_{\nu }{}^{c,\rho }$  &  ${g_{W}^3}/{180}$  \\
		\hline
		\multirow{5}{*}{$Q_H$} &  \multirow{5}{*}{$\left(H^{\dagger }H^{  })^3\right.$}  &  $-\frac{2}{15} \left(\left| \alpha_\eta\right| {}^6+\left| \alpha_\xi\right| {}^6\right)+\frac{2}{3} \left(\left| \beta_\eta\right| {}^6+\left| \beta_\xi\right| {}^6\right)  $  \\
		&&$+\frac{2}{3} \left(\left| \alpha_\eta\right| {}^4 \left| \beta_\eta\right| {}^2+\left| \alpha_\xi\right| {}^4 \left| \beta_\xi\right| {}^2\right)+2 \left(\left| \alpha_\eta\right| {}^2 \left| \beta_\eta\right| {}^4+\left| \alpha_\xi\right| {}^2 \left| \beta_\xi\right| {}^4\right)$\\
		&&$+\frac{2}{3} \left(\left| \alpha_\eta\right| {}^2 \left(\left(\alpha_\eta^*\right){}^2 \beta_\eta^2+\alpha_\eta^2 \left(\beta_\eta^*\right){}^2\right)+\left| \alpha_\xi\right| {}^2 \left(\left(\alpha_\xi^*\right){}^2 \beta_\xi^2+\alpha_\xi^2 \left(\beta_\xi^*\right){}^2\right)\right)$\\
		&&$+2 \left(\left| \beta_\eta\right| {}^2 \left(\left(\alpha_\eta^*\right)^2 \beta_\eta^2+\alpha_\eta^2 \left(\beta_\eta^*\right){}^2\right)+\left| \beta_\xi\right| {}^2 \left(\left(\alpha_\xi^*\right){}^2 \beta_\xi^2+\alpha_\xi^2 \left(\beta_\xi^*\right){}^2\right)\right)$\\
		&&$- 2 \lambda_H \mathcal{C}_{F} + \frac{4}{5} \lambda_H \left(\left|\alpha_\xi\right|^2+\left|\alpha_\eta \right|^2\right)+\frac{4}{3} \lambda_H \left(\left| \beta_\xi\right| {}^2+\left| \beta_\eta\right| {}^2\right)$\\
		\hline
		\multirow{3}{*}{$Q_{H\square }$}  &  \multirow{3}{*}{$\left(H^{\dagger }H^{  }\text{)$\square $(}H^{\dagger }H^{  }\right)$}  &  $-\frac{2}{5} \left(\left| \alpha_\eta\right| {}^2+\left| \alpha_\xi\right| {}^2\right){}^2-\frac{1}{3} \left(\left| \beta_\eta\right| {}^2+\left| \beta_\xi\right| {}^2\right){}^2$  \\
		&&$-\frac{1}{3} \left(\left| \beta_\xi\right| {}^2 \left| \alpha_\eta\right| {}^2+\left| \alpha_\xi\right| {}^2 \left| \beta_\eta\right| {}^2\right)-1 \left(\left| \alpha_\eta\right| {}^2 \left| \beta_\eta\right| {}^2+\left| \alpha_\xi\right| {}^2 \left| \beta_\xi\right| {}^2\right)$\\
		&&$-\frac{2}{3} \left(\alpha_\xi \beta_\xi^* \alpha_\eta^* \beta_\eta+\alpha_\xi^* \beta_\xi \alpha_\eta \beta_\eta^*\right)+\frac{1}{3} \left(\alpha_\eta^2 \left(\beta_\eta^*\right){}^2+\left(\alpha_\eta^*\right){}^2 \beta_\eta^2\right)$\\
		\hline
		\multirow{3}{*}{$Q_{{HD}}$}  &  \multirow{3}{*}{$\left(H^{\dagger }\mathcal{D}_{\mu }H^{  })^*\right(H^{\dagger }\mathcal{D}^{\mu }H^{  })$}  &  $-\frac{4}{5} \left(\left| \alpha_\xi\right| {}^2-\left| \alpha_\eta\right| {}^2\right){}^2-\frac{2}{3} \left(\left| \beta_\xi\right| {}^2-\left| \beta_\eta\right| {}^2\right){}^2$  \\
		&&$+\frac{2}{3} \left(\left| \beta_\xi\right| {}^2 \left| \alpha_\eta\right| {}^2+\left| \alpha_\xi\right| {}^2 \left| \beta_\eta\right| {}^2\right)-2 \left(\left| \alpha_\eta\right| {}^2 \left| \beta_\eta\right| {}^2+\left| \alpha_\xi\right| {}^2 \left| \beta_\xi\right| {}^2\right)$\\
		&&$+\frac{2}{3} \left(\alpha_\eta^2 \left(\beta_\eta^*\right){}^2+\left(\alpha_\eta^*\right){}^2 \beta_\eta^2\right)+\frac{4}{3} \left(\alpha_\xi \beta_\xi^* \alpha_\eta^* \beta_\eta+\alpha_\xi^* \beta_\xi \alpha_\eta \beta_\eta^*\right)$\\
		\hline
		\multirow{2}{*}{$Q_{{HB}}$}  &  \multirow{2}{*}{$\left(H^{\dagger }H^{  }\right)B_{\mu \nu }B^{\mu \nu }$}  &
		$\frac{g_{_Y}^2}{120}\left[(-7+40\mathcal{Y}-80\mathcal{Y}^2)|\alpha_\xi|^2+(-7-40\mathcal{Y}-80\mathcal{Y}^2)|\alpha_\eta|^2\right.$  \\
		&&$\left.+(5-40\mathcal{Y}+80\mathcal{Y}^2)|\beta_\xi|^2+(5+40\mathcal{Y}+80\mathcal{Y}^2)|\beta_\eta|^2\right]$\\
		\hline
		$Q_{{HW}}$  &  $\left(H^{\dagger }H^{  }\right)W_{\mu \nu }{}^aW^{a,\mu \nu }$  &  $-\frac{7 g_{_W}^2}{120}\left(|\alpha_\xi|^2+|\alpha_\eta|^2\right) + \frac{g_{_W}^2}{24}\left(|\beta_\xi|^2 + |\beta_\eta|^2\right)$  \\
		\hline
		\multirow{2}{*}{$Q_{{HWB}} $}  &  \multirow{2}{*}{$\left(H^{\dagger }\tau ^aH^{  }\right)W_{\mu \nu }{}^aB^{\mu \nu }$}  &  $\frac{g_{_W} g_{_Y}}{60}$$\left[(3-20\mathcal{Y})|\alpha_\xi|^2+(3+20\mathcal{Y})|\alpha_\eta|^2\right.$  \\
		&&$\left.+5(-1+4\mathcal{Y}) |\beta_\xi|^2 - 5 (1+4\mathcal{Y}) |\beta_\eta|^2\right]$\\
		\hline\hline
		$Q_{{eH}}$  &  $\left(H^{\dagger }H^{  }\right)\left(\bar{l}\text{ e }H^{  }\text{)+h.c.}\right.$  &  $-\frac{1}{2}\text{Re}\left[\left(Y^{e}_{_\text{SM}}\right)^{\dagger}\right] \mathcal{C}_F+\frac{1}{2}\text{Im}\left[\left(Y^{e}_{_\text{SM}}\right)^\dagger\right]\tilde{\mathcal{C}}_F + 2 \lambda_H \left(Y^{e}_{_\text{SM}}\right)^{\dagger}\left(Y^{e}_{_\text{SM}}\right) \mathcal{C}_{K4} $  \\
		\hline
		$Q_{{uH}}$  &  $\left(H^{\dagger }H^{  }\right)\left(\bar{q}\text{ u }\tilde{H}\right)\text{+h.c.}$  &  $-\frac{1}{2}\text{Re}\left[\left(Y^{u}_{_\text{SM}}\right)^{\dagger}\right] \mathcal{C}_F-\frac{1}{2}\text{Im}\left[\left(Y^{u}_{_\text{SM}}\right)^\dagger\right]\tilde{\mathcal{C}}_F+ 2 \lambda_H \left(Y^{u}_{_\text{SM}}\right)^{\dagger}\left(Y^{u}_{_\text{SM}}\right) \mathcal{C}_{K4}$  \\
		\hline
		$Q_{{dH}}$  &  $\left(H^{\dagger }H^{  }\right)\left(\bar{q}\text{ d }H^{  }\text{)+h.c.}\right.$  &  $-\frac{1}{2}\text{Re}\left[\left(Y^{d}_{_\text{SM}}\right)^{\dagger}\right] \mathcal{C}_F +\frac{1}{2}\text{Im}\left[\left(Y^{d}_{_\text{SM}}\right)^\dagger\right]\tilde{\mathcal{C}}_F + 2 \lambda_H \left(Y^{d}_{_\text{SM}}\right)^{\dagger}\left(Y^{d}_{_\text{SM}}\right) \mathcal{C}_{K4}$  \\
		\hline
		$Q_{{ledq}}$  &  $\left(\bar{l}^j\text{ e)(}\bar{d} q_j\text{)+h.c.}\right.$  &  $\left\{\left(Y^{e}_{_\text{SM}}\right)\left(Y^{d}_{_\text{SM}}\right)^\dagger \mathcal{C}_{K4}+\text{h.c.}\right\}$  \\
		\hline
		$Q_{{quqd}}^{(1)}$  &  $\left(\bar{q}^j\text{ u)}\epsilon _{\text{jk}}\right(\bar{q}^k\text{ d)+h.c.}$  &$\left\{\left(Y^{u}_{_\text{SM}}\right)^{\dagger}\left(Y^{d}_{_\text{SM}}\right)^\dagger \mathcal{C}_{K4}+\text{h.c.}\right\}$  \\
		\hline
		$Q_{{lequ}}^{(1)}$  &  $\left(\bar{l}^j\text{ e)}\epsilon _{\text{jk}}\right(\bar{q}^k\text{ u)+h.c.}$  & $-\left\{\left(Y^{e}_{_\text{SM}}\right)^{\dagger}\left(Y^{u}_{_\text{SM}}\right)^\dagger \mathcal{C}_{K4}+\text{h.c.}\right\}$  \\
		\hline
		$Q_{{le}}$  &  $\left(\bar{l} \gamma _{\mu }\text{ l)(}\bar{e} \gamma _{\mu }\text{ e)}\right.$  &$-\frac{1}{2}\left(Y^{e}_{_\text{SM}}\right)^{\dagger}\left(Y^{e}_{_\text{SM}}\right)\mathcal{C}_{K4}$  \\
		\hline
		$Q_{{qu}}^{(1)}$  &  $\left(\bar{q} \gamma ^{\mu }\text{ q)(}\bar{u} \gamma _{\mu }\text{ u)}\right.$  &$-\frac{1}{2}\left(Y^{u}_{_\text{SM}}\right)^{\dagger}\left(Y^{u}_{_\text{SM}}\right)\mathcal{C}_{K4}$  \\
		\hline
		$Q_{{qd}}^{(1)}$  &  $\left(\bar{q} \gamma _{\mu }\text{ q)(}\bar{d} \gamma _{\mu }\text{ d)}\right.$  &$-\frac{1}{2}\left(Y^{d}_{_\text{SM}}\right)^{\dagger}\left(Y^{d}_{_\text{SM}}\right)\mathcal{C}_{K4}$  \\
		\hline \hline
	\end{tabular}
	\end{threeparttable}
\end{table*}

Here, we define the following functions to express the WCs in much more compact form \cite{Angelescu:2020yzf}  in Tab.~\ref{tab:warsaw1}
\begin{align}
|\alpha_{i}|^2 = \frac{1}{4} \left( |Y_{i_L}|^2 + |Y_{i_R}|^2 + Y_{i_L}^* Y_{i_R} + Y_{i_L} Y_{i_R}^* \right),  \\
|\beta_{i}|^2 = \frac{1}{4} \left( |Y_{i_L}|^2 + |Y_{i_R}|^2 - Y_{i_L}^* Y_{i_R} - Y_{i_L} Y_{i_R}^* \right),
\end{align}
where, $i = \eta,\xi$. We further use the additional abbreviations for the same purpose \cite{Angelescu:2020yzf}
\begin{equation}
\begin{split}
\label{eq:cf}
\mathcal{C}_F = \ & -\frac{2}{5} \left(\left| \alpha_\xi\right| {}^4-4 \left| \alpha_\xi\right| {}^2 \left| \alpha_\eta\right| {}^2+\left| \alpha_\eta\right| {}^4\right)+\frac{4}{3} \left(\left| \beta_\eta\right| {}^4 + \left| \beta_\xi\right| {}^2 \left| \beta_\eta\right| {}^2 +\left| \beta_\xi\right| {}^4\right)\\
		&+2 \left(\left| \alpha_\eta\right| {}^2 \left| \beta_\eta\right| {}^2+\left| \alpha_\xi\right| {}^2 \left| \beta_\xi\right| {}^2\right)+\frac{2}{3} \left(\left| \beta_\xi\right| {}^2 \left| \alpha_\eta\right| {}^2+\left| \alpha_\xi\right| {}^2 \left| \beta_\eta\right| {}^2\right)\\
		&+\frac{4}{3} \left(\left(\alpha_\eta^*\right){}^2 \beta_\eta^2+\alpha_\eta^2 \left(\beta_\eta^*\right){}^2+\left(\alpha_\xi^*\right){}^2 \beta_\xi^2+\alpha_\xi^2 \left(\beta_\xi^*\right){}^2\right)+\frac{4}{3} \left(\alpha_\xi \beta_\xi^* \alpha_\eta^* \beta_\eta+\alpha_\xi^* \beta_\xi \alpha_\eta \beta_\eta^* \right)\,,\\
\end{split}
\end{equation}
\begin{equation*}
\begin{split}
		\mathcal{C}_{K4} = \ & \frac{1}{5}\left(\left|\alpha_\xi\right|^2+\left|\alpha_\eta \right|^2\right)+\frac{1}{3} \left(\left| \beta_\xi\right| {}^2+\left| \beta_\eta\right| {}^2\right)\,,\\
		\tilde{\mathcal{C}}_{F} = \ & \frac{1}{3}\left[\left(|Y_{\xi_{_L}}|^2+|Y_{\xi_{_R}}|^2\right)\text{Im}\left[Y_{\xi_{_L}}Y_{\xi_{_R}}^*\right]-\left(|Y_{\eta_{_L}}|^2+|Y_{\eta_{_R}}|^2\right)\text{Im}\left[Y_{\eta_{_L}}Y_{\eta_{_R}}^*\right]\right].\,
\end{split}
\end{equation*}
Here, we denote the electron ($e$)-, up ($u$)-, and down ($d$)-types Standard Model Yukawa couplings as $Y^{e}_{_\text{SM}}, Y^{u}_{_\text{SM}}  \text{ and }  Y^{d}_{_\text{SM}}$ respectively while we refer to the SM Higgs quartic self-coupling as $\lambda_H$.

The operators that may affect the couplings of gauge bosons to fermion currents, i.e., the relevant LHC processes are \cite{Grzadkowski:2010es, Brivio:2017btx,Dedes:2017zog} 
\begin{eqnarray}
\label{eq:oops}
&Q_{eB},Q_{eW},Q_{uB},Q_{uW}, Q_{dB},Q_{dW}, \\
&Q_{H l}^{(1)}, Q_{H l}^{(3)}, Q_{H q}^{(1)}, Q_{H q}^{(3)}, Q_{H u d}, Q_{H e}, Q_{H u}, Q_{H d}\,.
\end{eqnarray}
We have a relevant CP-even operator $Q_{HWB}$ that leads to an additional contribution to oblique corrections~\cite{Golden:1990ig,Holdom:1990tc,Altarelli:1990zd,Peskin:1990zt,Grinstein:1991cd,Altarelli:1991fk,Peskin:1991sw,Burgess:1993mg} and in particular the $S$ parameter. In later section, we discuss the impact of all relevant CP-even operators in Electro Weak Precision Observables (EWPOs) in detail. At this point it is worthy to mention that the operators
\begin{align}
\label{eq:oops2}
Q_{HB},Q_{HW},\\
Q_{H\widetilde W}, Q_{H\widetilde{B}},
\end{align}
do not modify trilinear gauge interactions as their contributions either vanish due to momentum conservation or can be absorbed into field and coupling redefinitions respecting gauge invariance. By investigating Tab.~\ref{tab:warsaw1}, we also find  that our adopted scenario, Eq.~\eqref{eq:vlike}, predicts ${\cal{C}}_{\widetilde{W}}=0$. Thus, together with our previous observations, we conclude that $Q_{H\widetilde{W}B}\neq 0$ is the only relevant operator to interpret  the results of ATLAS within the vector-like lepton framework.

In passing we would like to mention that some of the remaining non-zero operators can be probed in Higgs-boson associated final states or (to a lesser extent) through their radiative correction contributions~\cite{Grojean:2013kd,Englert:2014cva} (the latter corresponds to a two-loop suppression in the considered vector-like lepton UV completion). These processes provide additional CP sensitivity, however, at smaller Higgs-boson related production cross sections~(see e.g. the discussion in Ref.~\cite{Bernlochner:2018opw}) that receive corrections from a range of non-zero Wilson coefficients ${\cal{C}}_{H\widetilde{B}},{\cal{C}}_{H\widetilde{W}}$. We will not investigate Higgs-CP related effects in this work as neither they contribute to the electroweak precision observables nor impact the discussion of the previous section.\footnote{The additional chiral symmetry violation that leads to non-vanishing Wilson coefficients could in principle be traced into a uniform modification of the Higgs 2-point function~\cite{Englert:2019zmt} that can in principle be probed at hadron colliders. A related investigation was performed recently by CMS in four top final states~\cite{Sirunyan:2019wxt}. Sensitivity, however, is currently too limited for this effect to play an important role in a global fit.}

\begin{table*}[!b]
\parbox{0.5\textwidth}{
	\small
	\centering
	\begin{tabular}{|c|c|c|}
		\hline \hline
		Effective operators& Constrained& Constrained \\
		(Warsaw) & by EWPO & by Higgs-data \\
		\hline \hline
		$Q_H$&\cmark&\cmark\\
		\hline
		$Q_{H\square }$&\cmark&\cmark\\
		\hline
		$Q_{{HD}}$&\cmark&\cmark\\
		\hline
		$Q_{{HB}}$&\xmark&\cmark\\
		\hline
		$Q_{{HW}}$&\xmark&\cmark\\
		\hline
		$Q_{{HWB}}$&\cmark&\cmark\\
		\hline
		$Q_{{eH}}$&\xmark&\cmark\\
		\hline
		$Q_{{uH}}$&\xmark&\cmark\\
		\hline
		$Q_{{dH}}$&\xmark&\cmark\\
		\hline \hline
	\end{tabular}}
	\hfill
	\parbox{0.45\textwidth}{
	\vspace{-.3cm}\caption{\label{tab:effopsobs}The CP-even effective operators (in Warsaw basis) after integrating out VLLs: ``\cmark'' and ``\xmark'' signify that the respective operator is constrained or not respectively by the EWPOs and Higgs-data.  The operators $Q_{{ledq}},Q_{{quqd}}^{(1)}, Q_{{lequ}}^{(1)}, Q_{{le}}, Q_{{qu}}^{(1)}, Q_{{qd}}^{(1)}$ do not affect the observables under consideration.}}	
\end{table*}

\subsection{Constraints from Electroweak Precision Observables and Higgs-data}
\label{sec:elwco}

\begin{table*}[!b]
	\parbox{0.45\textwidth}{
		\vspace{-.1cm}
		\caption{\label{tab:fitparameters} Fitted values of the parameters, functions of Yukawa couplings of VLL model, using  EWPOs and the Higgs data. The choice of the parameters is guided by the detailed structures of the Wilson Coefficients, see Tab.~\ref{tab:warsaw1}. We assume $m=1~\text{TeV}$.}}
			\hfill
	\parbox{0.5\textwidth}{
		\small
		\centering
		\begin{tabular}{|c|c|}
			\hline \hline
			 VLLs:& \multirow{2}{*}{Fitted values of  parameters } \\
			Yukawa couplings& (@ 68\% C.L.)\\
			\hline \hline
			$\text{Re}\left[Y_{\eta_{_L}} Y_{\eta_{_R}}^\ast\right]$ & $1.00^{+6.50}_{-4.10}$ \\ \hline
			$\text{Im}\left[Y_{\eta_{_L}} Y_{\eta_{_R}}^\ast\right]$ & $0.07^{+2.77}_{-1.20}$ \\ \hline
			$\text{Re}\left[Y_{\xi_{_L}} Y_{\xi_{_R}}^\ast\right] $& $0.32^{+2.36}_{-4.51}$ \\ \hline
			$\text{Im}\left[Y_{\xi_{_L}} Y_{\xi_{_R}}^\ast \right]$& $-9.9^{+15.0}_{-22.8}$ \\ \hline
			$|Y_{\eta_{_L}}|^2$& $1.00^{+3.70}_{-1.00}$ \\ \hline
			$|Y_{\eta_{_R}}|^2$& $0.65^{+5.48}_{-0.65}$ \\ \hline
			$|Y_{\xi_{_L}}|^2$& $0.58^{+3.17}_{-0.58}$ \\ \hline
			$|Y_{\xi_{_R}}|^2$& $1.30^{+3.10}_{-1.30}$ \\ \hline
	\end{tabular}}
\end{table*}
The CP-even SMEFT operators contribute to the Electroweak Precision Observables (EWPOs)~\cite{Dawson_2020, Alonso:2013hga,Brivio:2017btx}, and to the production and decay of the SM Higgs~\cite{Murphy_2018}. We note that all the dimension six operators, generated after integrating out the VLLs, do not leave any impact to these observables, see Tab.~\ref{tab:effopsobs}.
We briefly outline the nature of correlations among the relevant effective operators and the EWPOs in the Appendix~\ref{app:ewpo}.   Though these observables do not constrain the CP-odd operators directly, we note that within our framework the CP-even and -odd  operators are related to each other,  see Tab.~\ref{tab:warsaw1}, through the model parameters, e.g., the Yukawa couplings in Eq.~(\ref{eq:vlike}). Here, we want to emphasize that instead of considering individual complex Yukawa couplings we prefer to work with a set of Yukawa-functions, chosen based on the computed WCs, see Tab.~\ref{tab:fitparameters}. This allows us to avoid unnecessary increase of free parameters in the theory which could have spoiled the quality of the fit without any gain for the earlier choice. Thus encapsulating the effects of these observables on CP-even WCs we can deduce  complementary constraints on the CP-odd WCs through the exotic Yukawa couplings in addition to the couplings' phases.
We perform a detail $\chi^2$-statistical analysis\footnote{We would like to mention that in our analysis the degree of freedom is 80 and $p$-value is $.36$. The min-$\chi^2$ is 83.86.} using a Mathematica package OptEx \cite{sunando_patra_2019_3404311} to estimate the allowed ranges of the model parameters in the light of  the  following experimental data: for EWPOs see Table~2 of Ref.~\cite{Baak_2014}, and Higgs data for Run-1 ATLAS and CMS \cite{Khachatryan:2016vau,Aad:2015gba} and Run-2 ATLAS and CMS \cite{Khachatryan:2016vau,Aad:2015gba,Aad:2019mbh,Aad:2020mkp,Aad:2020plj,ATLAS:2019ain,ATLAS:2020udg,Aaboud:2017jvq,Aaboud:2018urx,Aad:2019lpq,Sirunyan:2018koj,Sirunyan:2019qia}. 
The statistically estimated parameters which are suitably chosen functions of VLL-Yukawa couplings are depicted in Tab.~\ref{tab:fitparameters}.  

\begin{figure}[!t]
\begin{center}
   \subfigure{\includegraphics[width=0.48\textwidth]{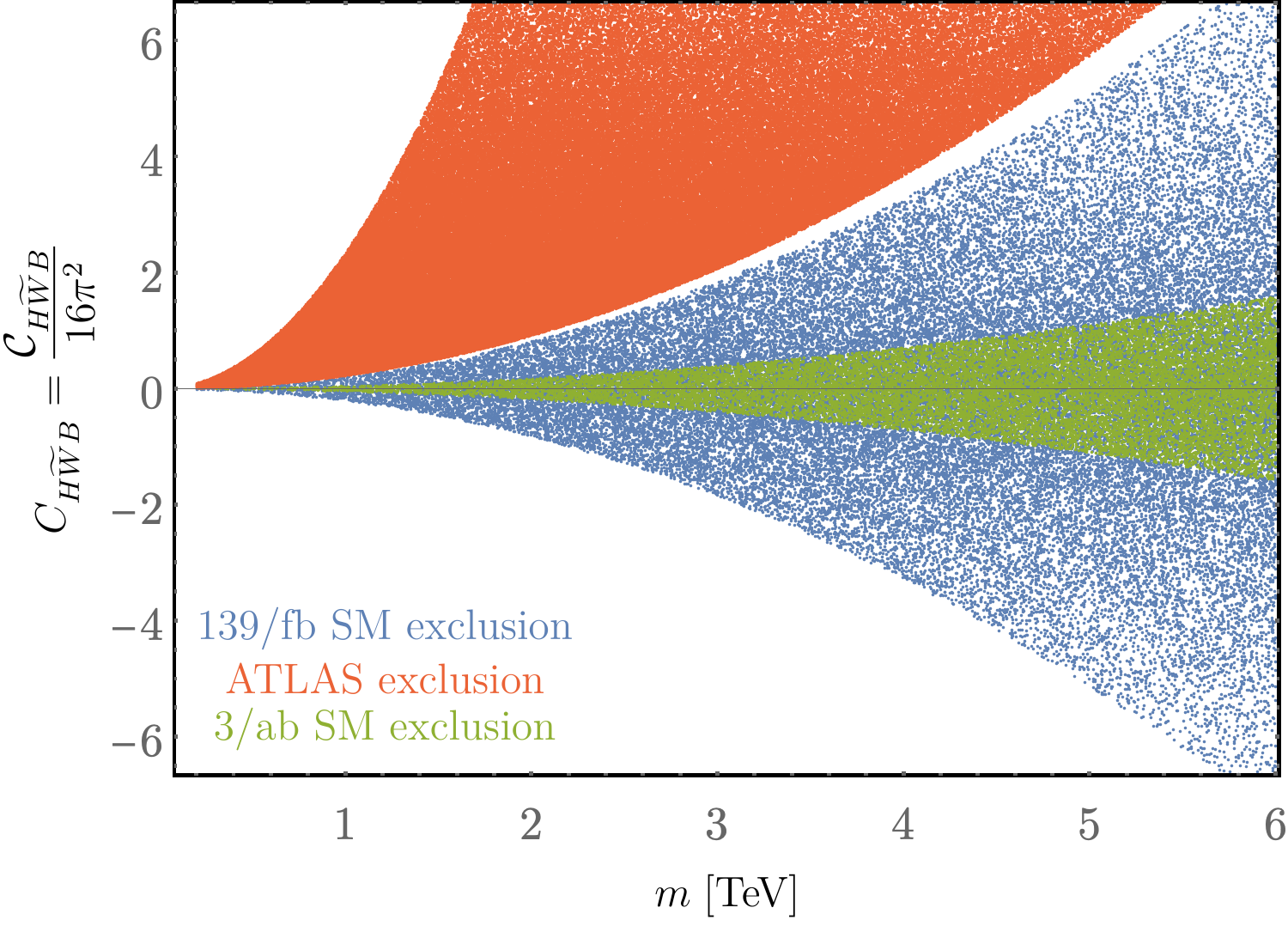}}
   \hskip 0.5cm
	\subfigure{\includegraphics[width=0.48\textwidth]{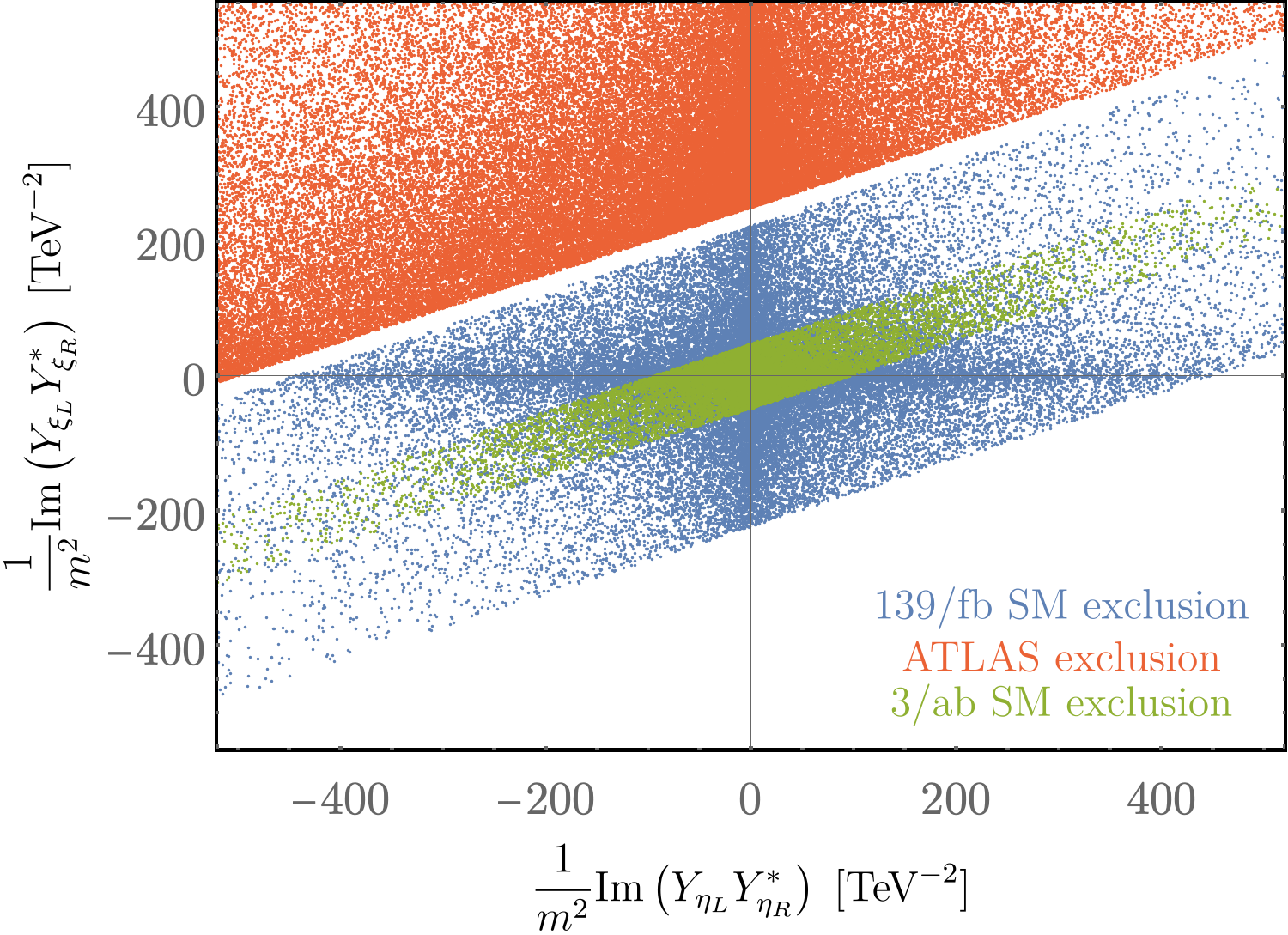}}
	\caption{On the left, $\chwb$ Wilson coefficient calculated from the expression in Tab.~\ref{tab:warsaw1} by sampling the Yukawa values and a fixed hypercharge of (SM-like) ${\cal{Y}}=-1/2$, is plotted against the vector-like lepton mass $m$. On the right the relevant combinations of Yukawa values contributing to $\chwb$ are shown. The points for the different exclusions are determined by assessing whether $\chwb/m^2$ lies within the $95$\% contours for $139/$fb and $3/$ab, as well as the allowed range the observed data of the ATLAS experiment~\cite{Aad:2020sle}. \label{fig:paramscans}}
\end{center}
\end{figure}

\section{Indirect Vector-like leptons: From Run-2 to the HL-LHC frontier}
\label{sec:extrapo}
In Fig.~\ref{fig:paramscans}, we show a scan over the model parameters when contrasted with the parameter constraints of the ATLAS analysis result of Eq.~\eqref{eq:tension}. It can be seen that that the large excess in the $C_{H\widetilde W B}$ 95\% constraint that is in tension with the SM favours either low mass scales or very large, potentially non-perturbative couplings. Direct searches for vector-like leptons have been discussed in \cite{Kumar:2015tna} and a HL-LHC direct coverage should be possible up to mass scales of 450 GeV which translates into model $\text{Im}(Y_{i_L}Y_{i_R}^\ast)\sim 40$ thus probing $\text{Re}(Y_i),\text{Im}(Y_i)\sim 6$. For such relatively low scales, where the EFT scale is identified with the statistical threshold of a particular analysis, the couplings are still in the strongly-coupled, yet perturbative $|Y|\lesssim 4\pi$ regime. Such large couplings, can lead to potential tension with other observables that are correlated through our particular model assumption. The constraints outlined in Sec.~\ref{sec:elwco} are in fact stronger, in particular for the combination of $\text{Im}(Y_{\xi_{_L}} Y_{\xi_{_R}}^\ast )\lesssim 40$.

\begin{figure}[!t]
\begin{center}
	\includegraphics[width=0.54\textwidth]{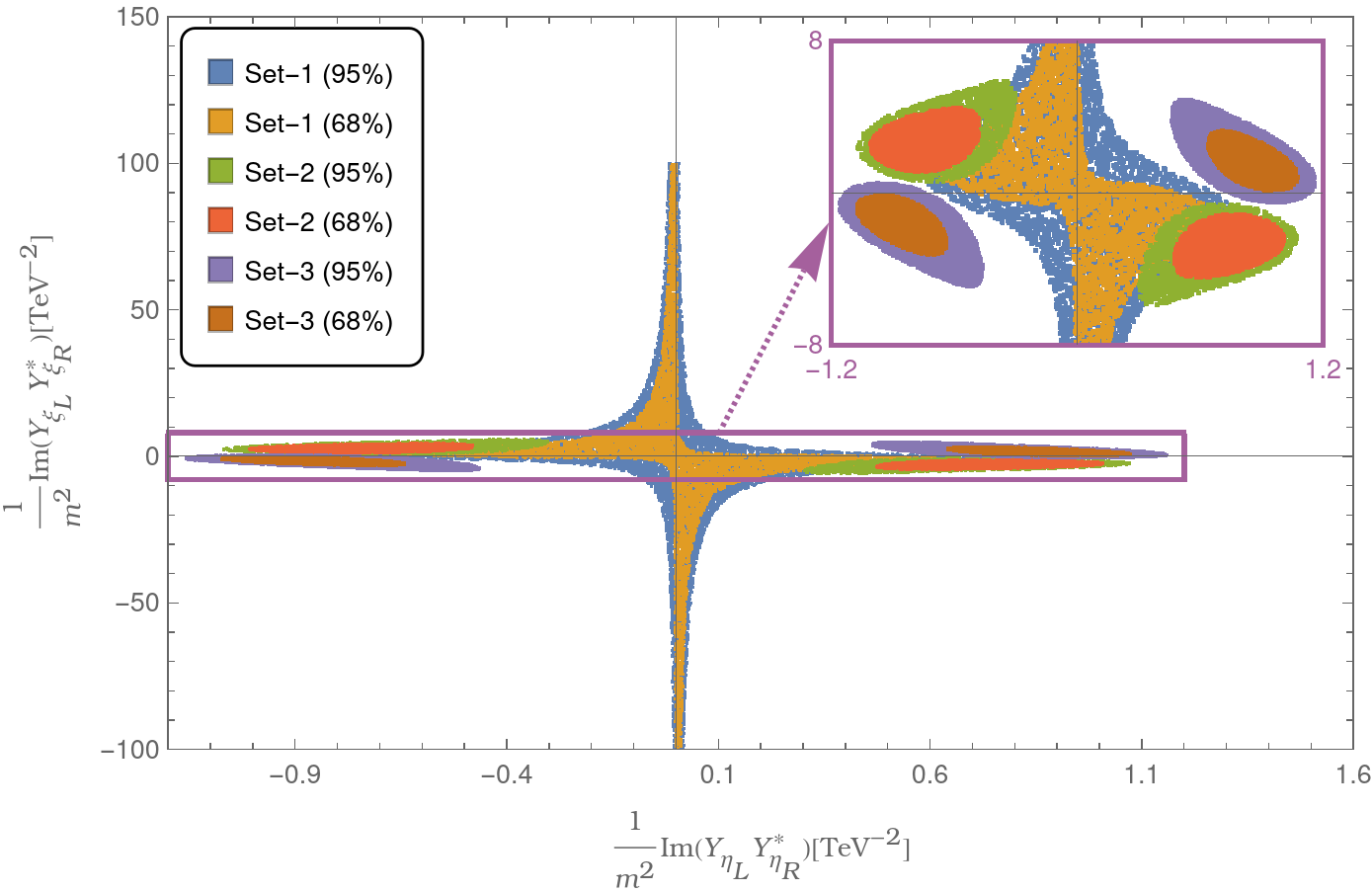}
	\includegraphics[width=0.44\textwidth]{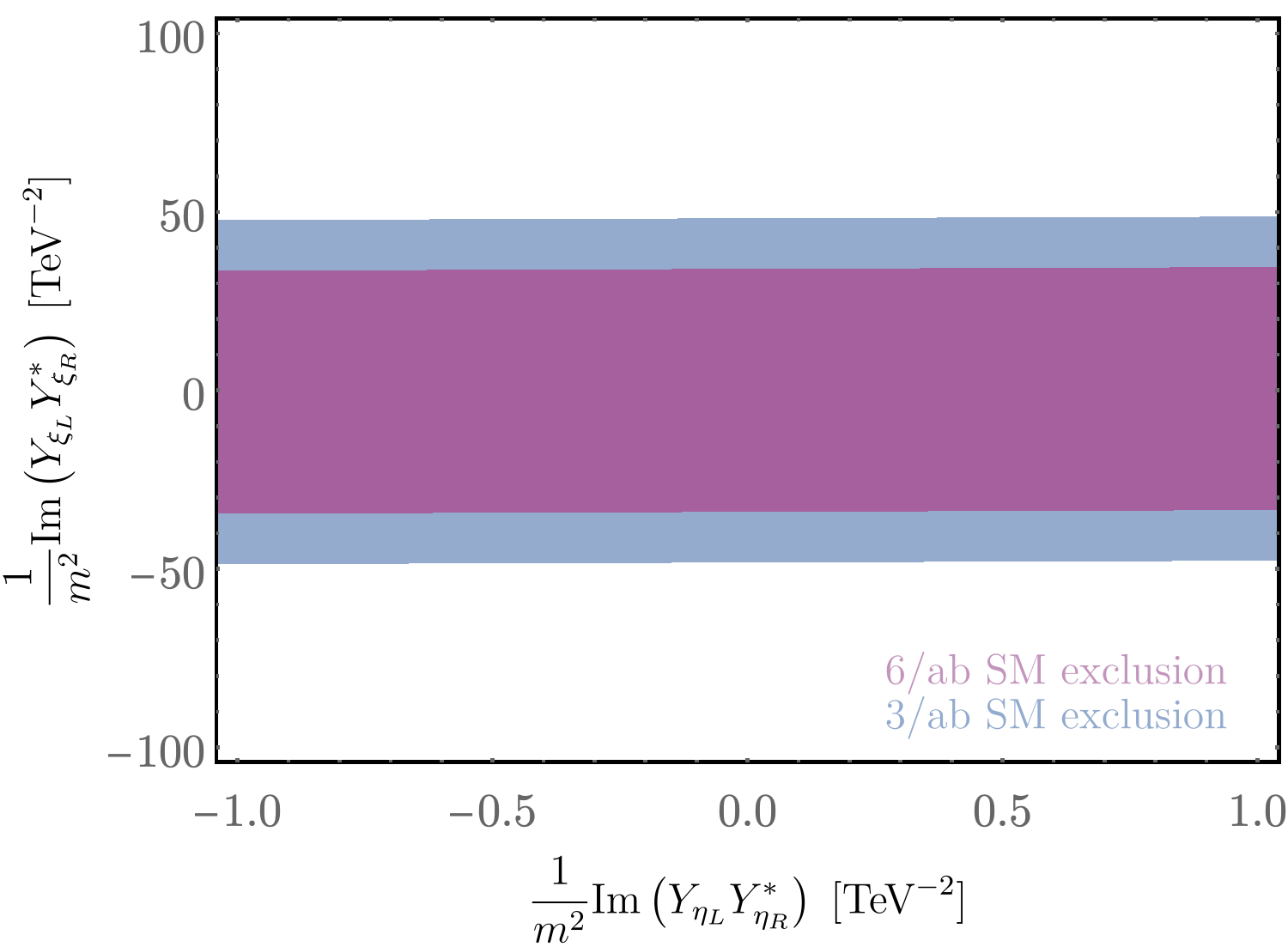}
   \caption{Yukawa values allowed by the $68\%$ and $95\%$~C.L. fits using EWPOs and SM Higgs decays plotted along with the regions allowed from the diboson analysis with $3/$ab. In addition, the diboson exclusions obtained for $6/$ab (resulting from a ATLAS+CMS combination) with the same methodology, are also included. {The $68\%$ and $95\%$~C.L. fits are shown for three benchmark points (see Tab.~\ref{tab:benchmark-points})}. \label{fig:ewpohiggs}}
\end{center}
\end{figure}

Returning to the complementary constraints that can be derived from the diboson, and in particular the $W\gamma$ analyses, we show the expected sensitivity range to the new physics scenario and its compatibility with the allowed parameter space consistent with the EWPO and Higgs signal strength in Fig.~\ref{fig:ewpohiggs}\footnote{
It is important to note that while generating the EWPO+Higgs data consistent parameter space, the CP violating observables are not included.}. {We choose three benchmark points (see Tab.~\ref{tab:benchmark-points}) and show the 65\% and 95\% C.L. regions in the $\text{Im}[Y_{\eta_L} Y_{\eta_R}^\ast]-\text{Im}[Y_{\xi_L} Y_{\xi_R}^\ast]$ plane}. There it becomes clear that the searches outlined in the beginning of this work will provide important sensitivity to this particular model class in the future in the $\text{Im}(Y_{\xi_{_L}} Y_{\xi_{_R}}^\ast)$ direction, which (for our choice of ${\cal{Y}}$) is relatively unconstrained by Higgs and EWPO data.

\begin{table*}[!b]
	\parbox{0.45\textwidth}{
		\vspace{-.1cm}
		\caption{ These are the three benchmark points chosen to analyze the 68\% and 95\% C.L.  allowed parameter space in the Im$[Y_{\eta_{_L}}Y_{\eta_{_R}}^\ast]$-Im$[ Y_{\xi_L} Y_{\xi_R}^\ast]$ plane, using the EWPO and the Higgs data. The shown fit parameters are set to the best-fit values (see Tab.~\ref{tab:warsaw1}) and two other set of points. The corresponding 68\% and 95\% C.L. regions are shown in Fig.~\ref{fig:ewpohiggs}.}}
	\parbox{0.5\textwidth}{
		\small
		\centering
	\begin{tabular}{|c|c|c|c|}
		\hline \hline
		\multirow{2}{*}{Fit parameters} & Set-1&\multirow{2}{*}{ \ \ Set-2 \ \ }& \multirow{2}{*}{\ \ Set-3 \ \ }\\
		&(Best fit)&&\\
		\hline \hline
		$\text{Re}\left[Y_{\eta_{_L}} Y_{\eta_{_R}}^\ast\right]$ & 1.00 &1.05& 0.25 \\ 
		\hline
		$\text{Re}\left[Y_{\xi_{_L}} Y_{\xi_{_R}}^\ast\right] $& 0.32 &0.19& 0.27\\
		\hline
		$|Y_{\eta_{_L}}|^2$& 1.00 &0.61& 0.73\\
		\hline
		$|Y_{\eta_{_R}}|^2$& 0.65 &1.2& 0.46\\
		\hline
		$|Y_{\xi_{_L}}|^2$& 0.58 &0.38& 0.52\\
		\hline
		$|Y_{\xi_{_R}}|^2$& 1.30 &0.05& 1.11\\
		\hline
	\end{tabular}\label{tab:benchmark-points}}
\end{table*}

\section{Discussion and Conclusions}
\label{sec:conc}
The insufficient amount CP violation in the SM to explain the observed matter-anti--matter asymmetry is a clear indication of the presence of new physics beyond the SM. Consequently analyses of CP properties of particle physics interactions are an important part of the current phenomenological program at various energies, reaching up to the current high-energy frontier explored at the LHC. The observation of $C_{H\widetilde W B}$-related excess by the ATLAS collaboration in the recent Ref.~\cite{Aad:2020sle} could be the first indication of the presence of such interactions in the gauge boson-Higgs sectors. Taking inspiration from Ref.~\cite{Aad:2020sle}, the focus of this work is two-fold:
\begin{enumerate}[(i)]
\item We motivate a particular UV model class, namely that of vector-like leptons Eq.~\eqref{eq:vlike}, to provide a minimal and consistent theoretical backdrop to the analysis of Ref.~\cite{Aad:2020sle}. We perform a complete matching calculation at one-loop order and demonstrate that all relevant CP-odd EFT deformations of the SM amplitude of diboson (and $Z+2j$) are dominantly captured by the $Q_{H\widetilde W B}$ operator. In parallel, at the given order we do not induce $Q_{\widetilde W}$, which impacts the analyses of CP-odd observables in (di)boson final states as well, but which is consistent with the SM expectation of $C_{\widetilde W}=0$ given the results of Ref.~\cite{Aad:2020sle}. The mass scales of the vector-like lepton scenario that can be directly explored at the LHC~\cite{Kumar:2015tna} constrain the model's parameters to the strong-coupling, yet perturbative regime. An analysis of electroweak precision and Run-2 Higgs results indicates that the region of the ATLAS excess could be explained by $\text{Im}(Y_{\eta_L}Y_{\eta_R}^\ast) \simeq 0$ and a significant $\text{Im}(Y_{\xi_L}Y_{\xi_R}^\ast)$ with some tension given the UV-model's correlation of CP-even and CP-odd couplings for masses that fall into the HL-LHC kinematic coverage. 
\item The excess observed by ATLAS deserves further scrutiny. We show that diboson analyses, and in particular $W\gamma$ production will serve as a strong cross check of the excess, in particular because its phenomenology is particularly sensitive to $Q_{H\widetilde W B}$-induced deviations. The analysis suggested in Sec.~\ref{sec:process}, will therefore allow the collaborations to directly explain the results of Ref.~\cite{Aad:2020sle} as a statistical fluctuation or gather further, strong evidence for a non-SM source of CP violation.
\end{enumerate}
Finally, the correlation of different Wilson coefficients as predicted by our matching calculation motivates additional Higgs-based phenomenology probe that can further constrain or solidify the excess through measurements that target, e.g. ${Q_{H\widetilde B}},{Q_{H\widetilde W}}$ in a suitable way~\cite{Bernlochner:2018opw} (see also~\cite{Huang:2020zde,Cirigliano:2019vfc}).

\acknowledgements
S.D.B. would like to thank Sunando Patra for the clarifications regarding the OptEx package and  Anisha for helpful discussions  on the statistical analysis.
The work of S.D.B. and J.C. is supported by the Science and Engineering Research Board,
Government of India, under the agreements SERB/PHY/2016348 (Early Career Research
Award) and SERB/PHY/2019501 (MATRICS).
C.E. is supported by the UK Science and Technology Facilities Council (STFC) under grants ST/P000746/1 and ST/T000945/1 and by the IPPP Associateship Scheme.
M.S. is supported by the STFC under grant ST/P001246/1. P.S. is supported by an STFC studentship under grant ST/T506102/1.

\pagebreak
\appendix

\section{Wilson Coefficients of the dimension six effective operators in the SILH-like basis}
\label{sec:silh}

\begin{table*}[!h]
	\caption{The complete set of most generic Wilson coefficients corresponding to the respective  dimension six effective operators in the SILH-like basis (a part of the result is noted in Ref.~\cite{Angelescu:2020yzf}). The heavy VLL multiplets are integrated out from the UV complete theory. The WCs are calculated up to one-loop order considering the CP conserving and violating couplings simultaneously. Note down the additional contributions of CPV couplings  to the CP-even operators. Note that we are referring to SILH-like operators as $O_i$, to highlight their difference from the Warsaw-basis operators.	\label{SILH}}
	\vskip 0.2cm
	\scriptsize
	\centering
	\renewcommand{\arraystretch}{1.2}
	\begin{tabular}{|c|c|l|}
		\hline \hline
		$\text{Operators}$&$\text{Operator Definition}$&$\text{Wilson coefficient} \ \left(\mathcal{C}_i\right)$\\
		\hline \hline
		$\tilde{O}_{f}$&$\frac{i}{2} |H|^2 \left(\left(\mathcal{D}^2H\right)^{\dagger} H -H^\dagger \mathcal{D}^2 H\right)$ &$\frac{1}{3}\left[\left(\left|Y_{\xi_L}\right|^2+\left|Y_{\xi_R}\right|^2\right)\text{Im}\left[Y_{\xi_L}Y_{\xi_R}^*\right]-\left(\left|Y_{\eta_L}\right|^2+\left|Y_{\eta_R}\right|^2\right)\text{Im}\left[Y_{\eta_L}Y_{\eta_R}^*\right]\right]$\\
		\hline
		$\tilde{O}_{BB}$  &$g_{_Y}^2\left(H^{\dagger }H^{  }\right)\tilde{B}_{\mu \nu }B^{\mu \nu }$  &  $-\frac{1}{12}\left[(1+6\mathcal{Y}+12\mathcal{Y}^2)\text{Im}[Y_{\eta_L} Y^*_{\eta_{R}}] +(1-6\mathcal{Y}+12\mathcal{Y}^2) \text{Im}[Y_{\xi_L} Y^*_{\xi_R}]\right]$  \\
		\hline
		$\tilde{O}_{WW}$  &$g_{_W}^2\left(H^{\dagger }H^{  }\right)\tilde{W}_{\mu \nu }{}^aW^{a,\mu \nu }$  &  $-\frac{1}{12}\text{Im}[Y_{\eta_L}Y^*_{\eta_R}+Y_{\xi_L}Y^*_{\xi_R}]$  \\
		\hline
		$\tilde{O}_{WB}$  & $2 g_{_W} g_{_Y}\left(H^{\dagger }\tau ^a H^{  }\right)\tilde{W}_{\mu \nu }{}^aB^{\mu \nu }$  &  $\frac{1}{12}\left[ (1+6\mathcal{Y}) \text{Im}[Y_{\eta_L} Y^*_{\eta_R}] + (1-6\mathcal{Y}) \text{Im}[Y_{\xi_L} Y^*_{\xi_R}] \right]$  \\
		\hline \hline
		$O_{3W}$  &  $\frac{g_{_W}^3}{3!}\epsilon ^{\text{abc}}W_{\rho }{}^{a,\mu }W_{\mu }{}^{b,\nu }W_{\nu }{}^{c,\rho }$  &  $\frac{1}{30}$  \\
		\hline
		$O_{2W}$  &  $-\frac{g_{_W}^2}{2}\left(\mathcal{D}^{\mu }W_{\mu \nu }{}^a)^2\right.$  &  $\frac{2}{15}$  \\
		\hline
		$O_{2B}$  &  $-\frac{g_{_Y}^2}{2}\left(\partial ^{\mu }B_{\mu \nu })^2\right.$  &  $\frac{2+16 \mathcal{Y}^2}{15}$  \\
		\hline
		$O_6$  &  $\left(H^{\dagger }H^{  })^3\right.$  &  $-\frac{2}{15} \left(\left| \alpha_{\eta}\right| {}^6+\left| \alpha_{\xi}\right| {}^6\right)+\frac{2}{3} \left(\left| \beta_{\eta}\right| {}^6+\left| \beta_{\xi}\right| {}^6\right) - 2 \lambda_H \mathcal{C}_{F} $  \\
		&&$+\frac{2}{3} \left(\left| \alpha_{\eta}\right| {}^4 \left| \beta_{\eta}\right| {}^2+\left| \alpha_{\xi}\right| {}^4 \left| \beta_{\xi}\right| {}^2\right)+2 \left(\left| \alpha_{\eta}\right| {}^2 \left| \beta_{\eta}\right| {}^4+\left| \alpha_{\xi}\right| {}^2 \left| \beta_{\xi}\right| {}^4\right)$\\
		&&$+\frac{2}{3} \left(\left| \alpha_{\eta}\right| {}^2 \left(\left(\alpha_{\eta}^*\right){}^2 \beta_{\eta}^2+\alpha_{\eta}^2 \left(\beta_{\eta}^*\right){}^2\right)+\left| \alpha_{\xi}\right| {}^2 \left(\left(\alpha_{\xi}^*\right){}^2 \beta_{\xi}^2+\alpha_{\xi}^2 \left(\beta_{\xi}^*\right){}^2\right)\right)$\\
		&&$+2 \left(\left| \beta_{\eta}\right| {}^2 \left(\left(\alpha_{\eta}^*\right)^2 \beta_{\eta}^2+\alpha_{\eta}^2 \left(\beta_{\eta}^*\right){}^2\right)+\left| \beta_{\xi}\right| {}^2 \left(\left(\alpha_{\xi}^*\right){}^2 \beta_{\xi}^2+\alpha_{\xi}^2 \left(\beta_{\xi}^*\right){}^2\right)\right)$  \\
		\hline
		$O_H$  &  $\frac{1}{2}\left(\partial _{\mu }\right(H^{\dagger }H^{  }))^2$  &  $\frac{4}{5} \left(\left| \alpha_{\eta}\right| {}^2+\left| \alpha_{\xi}\right| {}^2\right){}^2+\frac{2}{3} \left(\left| \beta_{\eta}\right| {}^2+\left| \beta_{\xi}\right| {}^2\right){}^2$ \\
		&&$+\frac{2}{3} \left(\left| \beta_{\xi}\right| {}^2 \left| \alpha_{\eta}\right| {}^2+\left| \alpha_{\xi}\right| {}^2 \left| \beta_{\eta}\right| {}^2\right)+2 \left(\left| \alpha_{\eta}\right| {}^2 \left| \beta_{\eta}\right| {}^2+\left| \alpha_{\xi}\right| {}^2 \left| \beta_{\xi}\right| {}^2\right)$\\
		&&$+\frac{4}{3} \left(\alpha_{\xi} \beta_{\xi}^* \alpha_{\eta}^* \beta_{\eta}+\alpha_{\xi}^* \beta_{\xi} \alpha_{\eta} \beta_{\eta}^*\right)-\frac{2}{3} \left(\alpha_{\eta}^2 \left(\beta_{\eta}^*\right){}^2+\left(\alpha_{\eta}^*\right){}^2 \beta_{\eta}^2\right)$\\
		\hline
		$O_T$  &  $\left|H^{\dagger }\mathcal{D}_{\mu }H^{  }\right|^2$  &  $-\frac{4}{5} \left(\left| \alpha_{\xi}\right| {}^2-\left| \alpha_{\eta}\right| {}^2\right){}^2-\frac{2}{3} \left(\left| \beta_{\xi}\right| {}^2-\left| \beta_{\eta}\right| {}^2\right){}^2$  \\
		&&$+\frac{2}{3} \left(\left| \beta_{\xi}\right| {}^2 \left| \alpha_{\eta}\right| {}^2+\left| \alpha_{\xi}\right| {}^2 \left| \beta_{\eta}\right| {}^2\right)-2 \left(\left| \alpha_{\eta}\right| {}^2 \left| \beta_{\eta}\right| {}^2+\left| \alpha_{\xi}\right| {}^2 \left| \beta_{\xi}\right| {}^2\right)$\\
		&&$+\frac{2}{3} \left(\alpha_{\eta}^2 \left(\beta_{\eta}^*\right){}^2+\left(\alpha_{\eta}^*\right){}^2 \beta_{\eta}^2\right)+\frac{4}{3} \left(\alpha_{\xi} \beta_{\xi}^* \alpha_{\eta}^* \beta_{\eta}+\alpha_{\xi}^* \beta_{\xi} \alpha_{\eta} \beta_{\eta}^*\right)$\\
		\hline
		$O_f$  &  $\frac{1}{2} |H|^2 \left(H^{\dagger }\mathcal{D}^2 H + h.c. )\right.$  &  $\mathcal{C}_F$\\
		\hline
		$O_{K4}$ & $|\mathcal{D}^2 H|^2$ & $\frac{1}{5}\left(\left|\alpha_\xi\right|^2+\left|\alpha_\eta\right|^2\right)+\frac{1}{3} \left(\left| \beta_{\xi}\right| {}^2+\left| \beta_{\eta}\right| {}^2\right)$\\
		\hline
		$O_{{BB}}$  &  $g_{_Y}^2\left(H^{\dagger }H^{  }\right)B_{\mu \nu }B^{\mu \nu }$  &  $\frac{1}{120}\left[(-7+40\mathcal{Y}-80\mathcal{Y}^2)|\alpha_{\xi}|^2+(-7-40\mathcal{Y}-80\mathcal{Y}^2)|\alpha_{\eta}|^2\right.$  \\
		&&$\left.+(5-40\mathcal{Y}+80\mathcal{Y}^2)|\beta_{\xi}|^2+(5+40\mathcal{Y}+80\mathcal{Y}^2)|\beta_{\eta}|^2\right]$  \\
		\hline
		$O_{{WB}}$  &  $2g_{_W}g_{_Y}\left(H^{\dagger }\tau ^aH^{  }\right)\left(W_{\mu \nu }{}^aB^{\mu \nu }\right)$  &  $\frac{1}{60}$$\left[(3-20\mathcal{Y})|\alpha_{\xi}|^2+(3+20\mathcal{Y})|\alpha_{\eta}|^2\right.$  \\
		&&$\left.+5(-1+4\mathcal{Y}) |\beta_{\xi}|^2 - 5 (1+4\mathcal{Y}) |\beta_{\eta}|^2\right]$  \\
		\hline
		$O_{{WW}}$  &  $g_{_W}^2\left(H^{\dagger }H^{  }\right)W_{\mu \nu }{}^aW^{a,\mu \nu }$  &  $-\frac{7 }{120}\left(|\alpha_{\xi}|^2+|\alpha_{\eta}|^2\right) + \frac{1}{24}\left(|\beta_{\xi}|^2 + |\beta_{\eta}|^2\right)$  \\
		\hline
		$O_W$  & $ i \, g_{_W}\left(H^{\dagger }\tau^a\overleftrightarrow{\mathcal{D}  }^{\mu }H^{  }\right)\left(\mathcal{D}^{\nu }W_{\mu \nu }{}^a\right)$  & $\frac{4}{15} \left(\left| \alpha_{\xi}\right| {}^2+\left| \alpha_{\eta}\right| {}^2\right) + \frac{1}{3}\left(\left|\beta_{\xi}\right|^2+\left|\beta_{\eta}\right|^2\right)$  \\
		\hline
		$O_B$  & $\frac{i}{2}g_{_Y}\left(H^{\dagger }\overleftrightarrow{\mathcal{D}  }^{\mu }H^{  }\right)\left(\partial ^{\nu }B_{\mu \nu }\right)$  & $\frac{4}{15} \left(\left| \alpha_{\xi}\right| {}^2+\left| \alpha_{\eta}\right| {}^2\right) + \frac{1}{3}\left(\left|\beta_{\xi}\right|^2+\left|\beta_{\eta}\right|^2\right)$  \\
		\hline
	\end{tabular}
\end{table*}

\section{Corrections to the EWPOs from dimension six Warsaw basis operators}
\label{app:ewpo}
The dimension six effective operators may affect the electroweak observables and modify the couplings related to the SM Higgs production and decay. These observables are very precisely measured. Thus any alteration beyond their SM predicted values puts stringent constraints on the WCs associated with those operators. 

The electroweak parameters under consideration are
\begin{eqnarray}\label{eq:SMparam}
\sin^2{\theta_{_W}}=\frac{1}{2}\left(1-\sqrt{1-\frac{4\pi\alpha}{\sqrt{2}G_{F} m^2_{Z}}}\right), & & 
g_{_Y}=\frac{\sqrt{4\pi\alpha}}{\cos\theta_{_{W}}},\;
g_{_W}=\frac{\sqrt{4\pi\alpha}}{\sin\theta_{_{W}}}, \nonumber\\	 
g_{_Z}=-\frac{g_{_W}}{\cos\theta_{_{W}}}, & & 
g^{SM}_{_L}=T_{3}- Q \sin^2\theta_{_{W}}, \,\,
g^{SM}_{_R}=- Q \sin^2\theta_{_{W}},\nonumber\\
<H>=v_{ew}=\frac{1}{2^{1/4}\sqrt{G_{F}}}, & & 
m^2_{_W}=m^2_{_Z}\cos^2 {\theta_{_{W}}},\nonumber
\end{eqnarray}
and can be expressed as functions of the electroweak input parameters fine structure constant $\alpha$,  mass of $Z$ boson  $m_{_Z}$,  and Fermi constant $G_{F}$\footnote{$G_F$ gets correction from $Q_{Hl}$ and $Q_{ll}$ dimension six operators. In the case of the model of Eq.~\eqref{eq:vlike} these two operators are absent, thus we directly impose $\delta G_F=0$.}.

Here,  we capture the additional contributions to the EWPOs, the relevant parameters and couplings, following the prescription suggested in Refs.~\cite{Dawson_2020, Alonso:2013hga, Brivio:2017btx}, in the presence of the computed dimension six operators in our VLL framework, see Tab.~\ref{tab:effopsobs}. 
We estimate the contributions to the following parameters based on Refs.~\cite{Dawson_2020, Alonso:2013hga,Brivio:2017btx} (we denote the $SU(2)_L, U(1)_Y$ gauge couplings with $g_{_W}$ and $g_{_Y}$, respectively):
\begin{itemize}
	\item $\alpha$ and $m_{_Z}$:
	\begin{eqnarray}
	\delta \alpha&=&\frac{ \alpha g_{_Y} g_{_W}  C_{{HWB}}}{(g_{_Y}^2+g_{_W}^2) \Lambda^2},\\
	\delta m^2_{_Z}&=&\frac{1}{2\sqrt{2}}\frac{m^2_{_Z}}{G_{F}}\frac{C_{{HD}}}{\Lambda^2}+\frac{2^{1/4}\sqrt{\pi \, \alpha} \, m_{_Z}}{G^{3/2}_{F}}\frac{C_{{HWB}}}{\Lambda^2},
	\end{eqnarray}
	 respectively.
	\item the Higgs boson mass $m_H$:
	\begin{eqnarray}
	\delta m^2_{H}&=&\frac{m^2_{H}}{\sqrt{2}G_{F}\Lambda^2}\left(-\frac{3C_{{H}}}{2\lambda_H}+2C_{{H}\square}-\frac{C_{{HD}}}{2}\right)\,,
	\end{eqnarray}

	\item the Weinberg angle ($\theta_{_{W}}$):
	\begin{eqnarray}
	\delta(\sin^2{\theta_{_{W}}})&=&\frac{\sin 2\theta_{_{W}}}{2\sqrt{2}\cos 2\theta_{_{W}}  G_{F}\Lambda^2}\left(\sin\theta_{_{W}} C_{HD}+2 C_{HWB}\right)\,,
	\end{eqnarray}

	\item the gauge coupling ($g_{_W}$):
	\begin{eqnarray}
	\delta g_{_W}&=&\frac{g_{_W}}{2}\left(2 \frac{\delta \alpha}{\alpha}-\frac{ \delta(\sin^2{\theta_{_{W}}})}{\sin^2{\theta_{_{W}}}}\right)\,,
	\end{eqnarray}
	
	\item the couplings of fermions to charged gauge bosons:
	\begin{eqnarray}\label{eq:fermigaugevert_charged}
	\delta(g^{l}_{_W})=\delta(g^{q}_{_W})=\delta g_{_W}\,,
	\end{eqnarray}
	
		\item the mass and width of $W$ boson:
	\begin{eqnarray*}
		\delta m_{_W}&=&\frac{m_{_Z} \cos \theta_{_{W}}}{2} \left(\frac{2\delta g_{_W}}{g_{_W}} \right)\,,\\
	\delta\Gamma_{W}&=&\Gamma_{W}\left(\frac{4}{3}\delta g^l_{W}+\frac{8}{3}\delta g^q_{W}+\frac{\delta m^2_{_W}}{2 m^2_{_W}}\right)\,,
	\end{eqnarray*}
	respectively,
	
	\item the couplings of left($L$) and right ($R$) chiral fermions to $Z$ boson:
	\begin{eqnarray}\label{eq:fermigaugevert_neutral}
	\frac{\delta g_{_Z}}{g_{_Z}}= -\frac{\delta m^2_{Z}}{2 m^2_{Z}}+\frac{\sin\theta_{_{W}} \cos\theta_{_{W}}}{\sqrt{2}G_{F}\Lambda^2}C_{{HWB}}\,, & & \nonumber\\ 
	\delta g^l_{L}=\delta(g_{_Z}) g^{l}_{L} + g_{_Z}\,\delta(\sin^{2}\theta_{_{W}})\,,  
	\delta g^{\nu}_{L} =\delta(g_{_Z}) g^{\nu}_{L}\,, &  &
		\delta g^l_{R} =\delta(g_{_Z}) g^{l}_{R}\,, 
		\delta g^{\nu}_{R} =	0\,, \nonumber \\
	\delta g^u_{L} =\delta(g_{_Z}) g^{u}_{L}+\frac{2}{3}g_{_Z} \,\delta(\sin^{2}\theta_{_{W}})\,, & & 	\delta g^u_{R} =\delta(g_{_Z})
	g^{u}_{R}+\frac{2}{3}g_{_Z}\,\delta(\sin^{2}\theta_{_{W}})\,, \nonumber \\
	\delta g^d_{L} =\delta(g_{_Z})	g^{d}_{L}+\frac{1}{3}g_{_Z}\,\delta(\sin^{2}\theta_{_{W}})\,, & & 
		\delta g^d_{R}  =\delta(g_{_Z})	g^{d}_{R}\,. \nonumber
	\end{eqnarray}
\end{itemize}

The total scattering cross section of $Z$ boson
\begin{eqnarray}
\sigma^{0}_{had}=\frac{12\pi}{m_{_Z}^2} \frac{\Gamma_{e}\Gamma_{had}}{\Gamma_{Z}^2}\,,\nonumber
\end{eqnarray}
including the effects of  $\delta \Gamma_{Z}$, $\delta \Gamma_{e}$ and $\delta \Gamma_{had}$, $\delta \sigma_{had}$ can then be calculated straightforwardly. The partial decay width of the $Z$ boson into fermions is given by
\begin{eqnarray}
\Gamma_{f}=N_{c}\frac{m_{_Z}}{12\pi}\sqrt{1-4\frac{m_{f}^2}{m_{_Z}^2}}\left(\frac{1}{2}(g_{L}^2+g_{R}^2)+\frac{2m_{f}^2}{m_{_Z}^2}\Big(-\frac{g_{L}^2}{4}-\frac{g_{R}^2}{4}-\frac{3}{2}g_{L} g_{R}\Big)\right),
\end{eqnarray}	
where $N_C$ is the color charge of the fermions. The change in the partial decay width is computed in a very similar way as done for $\delta\Gamma_{W}$. Furthermore, the ratios of the 
changes in partial decays, e.g.,  $\delta R_{l}$, $\delta R_{b}$ and $\delta R_{c}$, and the asymmetries $(\delta A_{f})$ and forward-backward  $(\delta A^{f}_{FB})$  can be recast in terms of changes of the couplings.

\bibliography{references} 

\begin{thebibliography}{73}
\expandafter\ifx\csname natexlab\endcsname\relax\def\natexlab#1{#1}\fi
\expandafter\ifx\csname bibnamefont\endcsname\relax
  \def\bibnamefont#1{#1}\fi
\expandafter\ifx\csname bibfnamefont\endcsname\relax
  \def\bibfnamefont#1{#1}\fi
\expandafter\ifx\csname citenamefont\endcsname\relax
  \def\citenamefont#1{#1}\fi
\expandafter\ifx\csname url\endcsname\relax
  \def\url#1{\texttt{#1}}\fi
\expandafter\ifx\csname urlprefix\endcsname\relax\def\urlprefix{URL }\fi
\providecommand{\bibinfo}[2]{#2}
\providecommand{\eprint}[2][]{\url{#2}}

\bibitem[{\citenamefont{Sakharov}(1967)}]{Sakharov:1967dj}
\bibinfo{author}{\bibfnamefont{A.~D.} \bibnamefont{Sakharov}},
  \bibinfo{journal}{Pisma Zh. Eksp. Teor. Fiz.} \textbf{\bibinfo{volume}{5}},
  \bibinfo{pages}{32} (\bibinfo{year}{1967}), \bibinfo{note}{[JETP
  Lett.5,24(1967); Sov. Phys. Usp.34,no.5,392(1991); Usp. Fiz.
  Nauk161,no.5,61(1991)]}.

\bibitem[{\citenamefont{Aad et~al.}(2020{\natexlab{a}})}]{Aad:2020ivc}
\bibinfo{author}{\bibfnamefont{G.}~\bibnamefont{Aad}} \bibnamefont{et~al.}
  (\bibinfo{collaboration}{ATLAS}), \bibinfo{journal}{Phys. Rev. Lett.}
  \textbf{\bibinfo{volume}{125}}, \bibinfo{pages}{061802}
  (\bibinfo{year}{2020}{\natexlab{a}}), \eprint{2004.04545}.

\bibitem[{\citenamefont{Sirunyan
  et~al.}(2020{\natexlab{a}})}]{Sirunyan:2020sum}
\bibinfo{author}{\bibfnamefont{A.~M.} \bibnamefont{Sirunyan}}
  \bibnamefont{et~al.} (\bibinfo{collaboration}{CMS}), \bibinfo{journal}{Phys.
  Rev. Lett.} \textbf{\bibinfo{volume}{125}}, \bibinfo{pages}{061801}
  (\bibinfo{year}{2020}{\natexlab{a}}), \eprint{2003.10866}.

\bibitem[{\citenamefont{Aad et~al.}(2020{\natexlab{b}})}]{Aad:2020sle}
\bibinfo{author}{\bibfnamefont{G.}~\bibnamefont{Aad}} \bibnamefont{et~al.}
  (\bibinfo{collaboration}{ATLAS}) (\bibinfo{year}{2020}{\natexlab{b}}),
  \eprint{2006.15458}.

\bibitem[{\citenamefont{Grzadkowski et~al.}(2010)\citenamefont{Grzadkowski,
  Iskrzynski, Misiak, and Rosiek}}]{Grzadkowski:2010es}
\bibinfo{author}{\bibfnamefont{B.}~\bibnamefont{Grzadkowski}},
  \bibinfo{author}{\bibfnamefont{M.}~\bibnamefont{Iskrzynski}},
  \bibinfo{author}{\bibfnamefont{M.}~\bibnamefont{Misiak}}, \bibnamefont{and}
  \bibinfo{author}{\bibfnamefont{J.}~\bibnamefont{Rosiek}},
  \bibinfo{journal}{JHEP} \textbf{\bibinfo{volume}{10}}, \bibinfo{pages}{085}
  (\bibinfo{year}{2010}), \eprint{1008.4884}.

\bibitem[{\citenamefont{Arnold et~al.}(2009)}]{Arnold:2008rz}
\bibinfo{author}{\bibfnamefont{K.}~\bibnamefont{Arnold}} \bibnamefont{et~al.},
  \bibinfo{journal}{Comput. Phys. Commun.} \textbf{\bibinfo{volume}{180}},
  \bibinfo{pages}{1661} (\bibinfo{year}{2009}), \eprint{0811.4559}.

\bibitem[{\citenamefont{Baglio et~al.}(2014)}]{Baglio:2014uba}
\bibinfo{author}{\bibfnamefont{J.}~\bibnamefont{Baglio}} \bibnamefont{et~al.}
  (\bibinfo{year}{2014}), \eprint{1404.3940}.

\bibitem[{\citenamefont{Bellm et~al.}(2016)}]{Bellm:2015jjp}
\bibinfo{author}{\bibfnamefont{J.}~\bibnamefont{Bellm}} \bibnamefont{et~al.},
  \bibinfo{journal}{Eur. Phys. J.} \textbf{\bibinfo{volume}{C76}},
  \bibinfo{pages}{196} (\bibinfo{year}{2016}), \eprint{1512.01178}.

\bibitem[{\citenamefont{Aad et~al.}(2011)}]{Aad:2011tc}
\bibinfo{author}{\bibfnamefont{G.}~\bibnamefont{Aad}} \bibnamefont{et~al.}
  (\bibinfo{collaboration}{ATLAS}), \bibinfo{journal}{JHEP}
  \textbf{\bibinfo{volume}{09}}, \bibinfo{pages}{072} (\bibinfo{year}{2011}),
  \eprint{1106.1592}.

\bibitem[{\citenamefont{Chatrchyan et~al.}(2014)}]{Chatrchyan:2013fya}
\bibinfo{author}{\bibfnamefont{S.}~\bibnamefont{Chatrchyan}}
  \bibnamefont{et~al.} (\bibinfo{collaboration}{CMS}), \bibinfo{journal}{Phys.
  Rev.} \textbf{\bibinfo{volume}{D89}}, \bibinfo{pages}{092005}
  (\bibinfo{year}{2014}), \eprint{1308.6832}.

\bibitem[{\citenamefont{Goebel et~al.}(1981)\citenamefont{Goebel, Halzen, and
  Leveille}}]{Goebel:1980es}
\bibinfo{author}{\bibfnamefont{C.~J.} \bibnamefont{Goebel}},
  \bibinfo{author}{\bibfnamefont{F.}~\bibnamefont{Halzen}}, \bibnamefont{and}
  \bibinfo{author}{\bibfnamefont{J.~P.} \bibnamefont{Leveille}},
  \bibinfo{journal}{Phys. Rev.} \textbf{\bibinfo{volume}{D23}},
  \bibinfo{pages}{2682} (\bibinfo{year}{1981}).

\bibitem[{\citenamefont{Brodsky and Brown}(1982)}]{Brodsky:1982sh}
\bibinfo{author}{\bibfnamefont{S.~J.} \bibnamefont{Brodsky}} \bibnamefont{and}
  \bibinfo{author}{\bibfnamefont{R.~W.} \bibnamefont{Brown}},
  \bibinfo{journal}{Phys. Rev. Lett.} \textbf{\bibinfo{volume}{49}},
  \bibinfo{pages}{966} (\bibinfo{year}{1982}).

\bibitem[{\citenamefont{Brown et~al.}(1983)\citenamefont{Brown, Kowalski, and
  Brodsky}}]{Brown:1982xx}
\bibinfo{author}{\bibfnamefont{R.~W.} \bibnamefont{Brown}},
  \bibinfo{author}{\bibfnamefont{K.~L.} \bibnamefont{Kowalski}},
  \bibnamefont{and} \bibinfo{author}{\bibfnamefont{S.~J.}
  \bibnamefont{Brodsky}}, \bibinfo{journal}{Phys. Rev.}
  \textbf{\bibinfo{volume}{D28}}, \bibinfo{pages}{624} (\bibinfo{year}{1983}),
  \bibinfo{note}{[Addendum: Phys. Rev.D29,2100(1984)]}.

\bibitem[{\citenamefont{Baur et~al.}(1993)\citenamefont{Baur, Han, and
  Ohnemus}}]{Baur:1993ir}
\bibinfo{author}{\bibfnamefont{U.}~\bibnamefont{Baur}},
  \bibinfo{author}{\bibfnamefont{T.}~\bibnamefont{Han}}, \bibnamefont{and}
  \bibinfo{author}{\bibfnamefont{J.}~\bibnamefont{Ohnemus}},
  \bibinfo{journal}{Phys. Rev. D} \textbf{\bibinfo{volume}{48}},
  \bibinfo{pages}{5140} (\bibinfo{year}{1993}), \eprint{hep-ph/9305314}.

\bibitem[{\citenamefont{Baur et~al.}(1994)\citenamefont{Baur, Errede, and
  Landsberg}}]{Baur:1994sa}
\bibinfo{author}{\bibfnamefont{U.}~\bibnamefont{Baur}},
  \bibinfo{author}{\bibfnamefont{S.}~\bibnamefont{Errede}}, \bibnamefont{and}
  \bibinfo{author}{\bibfnamefont{G.~L.} \bibnamefont{Landsberg}},
  \bibinfo{journal}{Phys. Rev.} \textbf{\bibinfo{volume}{D50}},
  \bibinfo{pages}{1917} (\bibinfo{year}{1994}), \eprint{hep-ph/9402282}.

\bibitem[{\citenamefont{Han}(1995)}]{Han:1995ef}
\bibinfo{author}{\bibfnamefont{T.}~\bibnamefont{Han}}, \bibinfo{journal}{AIP
  Conf. Proc.} \textbf{\bibinfo{volume}{350}}, \bibinfo{pages}{224}
  (\bibinfo{year}{1995}), \eprint{hep-ph/9506286}.

\bibitem[{\citenamefont{Aihara et~al.}(1995)}]{Aihara:1995iq}
\bibinfo{author}{\bibfnamefont{H.}~\bibnamefont{Aihara}} \bibnamefont{et~al.},
  pp. \bibinfo{pages}{488--546} (\bibinfo{year}{1995}),
  \eprint{hep-ph/9503425}.

\bibitem[{\citenamefont{Christensen and Duhr}(2009)}]{Christensen:2008py}
\bibinfo{author}{\bibfnamefont{N.~D.} \bibnamefont{Christensen}}
  \bibnamefont{and} \bibinfo{author}{\bibfnamefont{C.}~\bibnamefont{Duhr}},
  \bibinfo{journal}{Comput. Phys. Commun.} \textbf{\bibinfo{volume}{180}},
  \bibinfo{pages}{1614} (\bibinfo{year}{2009}), \eprint{0806.4194}.

\bibitem[{\citenamefont{Alloul et~al.}(2014)\citenamefont{Alloul, Christensen,
  Degrande, Duhr, and Fuks}}]{Alloul:2013bka}
\bibinfo{author}{\bibfnamefont{A.}~\bibnamefont{Alloul}},
  \bibinfo{author}{\bibfnamefont{N.~D.} \bibnamefont{Christensen}},
  \bibinfo{author}{\bibfnamefont{C.}~\bibnamefont{Degrande}},
  \bibinfo{author}{\bibfnamefont{C.}~\bibnamefont{Duhr}}, \bibnamefont{and}
  \bibinfo{author}{\bibfnamefont{B.}~\bibnamefont{Fuks}},
  \bibinfo{journal}{Comput. Phys. Commun.} \textbf{\bibinfo{volume}{185}},
  \bibinfo{pages}{2250} (\bibinfo{year}{2014}), \eprint{1310.1921}.

\bibitem[{\citenamefont{Degrande et~al.}(2012)\citenamefont{Degrande, Duhr,
  Fuks, Grellscheid, Mattelaer, and Reiter}}]{Degrande:2011ua}
\bibinfo{author}{\bibfnamefont{C.}~\bibnamefont{Degrande}},
  \bibinfo{author}{\bibfnamefont{C.}~\bibnamefont{Duhr}},
  \bibinfo{author}{\bibfnamefont{B.}~\bibnamefont{Fuks}},
  \bibinfo{author}{\bibfnamefont{D.}~\bibnamefont{Grellscheid}},
  \bibinfo{author}{\bibfnamefont{O.}~\bibnamefont{Mattelaer}},
  \bibnamefont{and} \bibinfo{author}{\bibfnamefont{T.}~\bibnamefont{Reiter}},
  \bibinfo{journal}{Comput. Phys. Commun.} \textbf{\bibinfo{volume}{183}},
  \bibinfo{pages}{1201} (\bibinfo{year}{2012}), \eprint{1108.2040}.

\bibitem[{\citenamefont{Alwall et~al.}(2011)\citenamefont{Alwall, Herquet,
  Maltoni, Mattelaer, and Stelzer}}]{Alwall:2011uj}
\bibinfo{author}{\bibfnamefont{J.}~\bibnamefont{Alwall}},
  \bibinfo{author}{\bibfnamefont{M.}~\bibnamefont{Herquet}},
  \bibinfo{author}{\bibfnamefont{F.}~\bibnamefont{Maltoni}},
  \bibinfo{author}{\bibfnamefont{O.}~\bibnamefont{Mattelaer}},
  \bibnamefont{and} \bibinfo{author}{\bibfnamefont{T.}~\bibnamefont{Stelzer}},
  \bibinfo{journal}{JHEP} \textbf{\bibinfo{volume}{06}}, \bibinfo{pages}{128}
  (\bibinfo{year}{2011}), \eprint{1106.0522}.

\bibitem[{\citenamefont{de~Aquino et~al.}(2012)\citenamefont{de~Aquino, Link,
  Maltoni, Mattelaer, and Stelzer}}]{deAquino:2011ub}
\bibinfo{author}{\bibfnamefont{P.}~\bibnamefont{de~Aquino}},
  \bibinfo{author}{\bibfnamefont{W.}~\bibnamefont{Link}},
  \bibinfo{author}{\bibfnamefont{F.}~\bibnamefont{Maltoni}},
  \bibinfo{author}{\bibfnamefont{O.}~\bibnamefont{Mattelaer}},
  \bibnamefont{and} \bibinfo{author}{\bibfnamefont{T.}~\bibnamefont{Stelzer}},
  \bibinfo{journal}{Comput. Phys. Commun.} \textbf{\bibinfo{volume}{183}},
  \bibinfo{pages}{2254} (\bibinfo{year}{2012}), \eprint{1108.2041}.

\bibitem[{\citenamefont{Alwall et~al.}(2014)\citenamefont{Alwall, Frederix,
  Frixione, Hirschi, Maltoni, Mattelaer, Shao, Stelzer, Torrielli, and
  Zaro}}]{Alwall:2014hca}
\bibinfo{author}{\bibfnamefont{J.}~\bibnamefont{Alwall}},
  \bibinfo{author}{\bibfnamefont{R.}~\bibnamefont{Frederix}},
  \bibinfo{author}{\bibfnamefont{S.}~\bibnamefont{Frixione}},
  \bibinfo{author}{\bibfnamefont{V.}~\bibnamefont{Hirschi}},
  \bibinfo{author}{\bibfnamefont{F.}~\bibnamefont{Maltoni}},
  \bibinfo{author}{\bibfnamefont{O.}~\bibnamefont{Mattelaer}},
  \bibinfo{author}{\bibfnamefont{H.~S.} \bibnamefont{Shao}},
  \bibinfo{author}{\bibfnamefont{T.}~\bibnamefont{Stelzer}},
  \bibinfo{author}{\bibfnamefont{P.}~\bibnamefont{Torrielli}},
  \bibnamefont{and} \bibinfo{author}{\bibfnamefont{M.}~\bibnamefont{Zaro}},
  \bibinfo{journal}{JHEP} \textbf{\bibinfo{volume}{07}}, \bibinfo{pages}{079}
  (\bibinfo{year}{2014}), \eprint{1405.0301}.

\bibitem[{\citenamefont{Alwall et~al.}(2007)}]{Alwall:2006yp}
\bibinfo{author}{\bibfnamefont{J.}~\bibnamefont{Alwall}} \bibnamefont{et~al.},
  \bibinfo{journal}{Comput. Phys. Commun.} \textbf{\bibinfo{volume}{176}},
  \bibinfo{pages}{300} (\bibinfo{year}{2007}), \eprint{hep-ph/0609017}.

\bibitem[{\citenamefont{Khachatryan et~al.}(2017)}]{Khachatryan:2016tgp}
\bibinfo{author}{\bibfnamefont{V.}~\bibnamefont{Khachatryan}}
  \bibnamefont{et~al.} (\bibinfo{collaboration}{CMS}), \bibinfo{journal}{Phys.
  Lett.} \textbf{\bibinfo{volume}{B766}}, \bibinfo{pages}{268}
  (\bibinfo{year}{2017}), \eprint{1607.06943}.

\bibitem[{\citenamefont{Aaboud et~al.}(2019)}]{Aaboud:2019nkz}
\bibinfo{author}{\bibfnamefont{M.}~\bibnamefont{Aaboud}} \bibnamefont{et~al.}
  (\bibinfo{collaboration}{ATLAS}), \bibinfo{journal}{Eur. Phys. J.}
  \textbf{\bibinfo{volume}{C79}}, \bibinfo{pages}{884} (\bibinfo{year}{2019}),
  \eprint{1905.04242}.

\bibitem[{\citenamefont{Campbell and Ellis}(1999)}]{Campbell:1999ah}
\bibinfo{author}{\bibfnamefont{J.~M.} \bibnamefont{Campbell}} \bibnamefont{and}
  \bibinfo{author}{\bibfnamefont{R.~K.} \bibnamefont{Ellis}},
  \bibinfo{journal}{Phys. Rev.} \textbf{\bibinfo{volume}{D60}},
  \bibinfo{pages}{113006} (\bibinfo{year}{1999}), \eprint{hep-ph/9905386}.

\bibitem[{\citenamefont{Campbell et~al.}(2011)\citenamefont{Campbell, Ellis,
  and Williams}}]{Campbell:2011bn}
\bibinfo{author}{\bibfnamefont{J.~M.} \bibnamefont{Campbell}},
  \bibinfo{author}{\bibfnamefont{R.~K.} \bibnamefont{Ellis}}, \bibnamefont{and}
  \bibinfo{author}{\bibfnamefont{C.}~\bibnamefont{Williams}},
  \bibinfo{journal}{JHEP} \textbf{\bibinfo{volume}{07}}, \bibinfo{pages}{018}
  (\bibinfo{year}{2011}), \eprint{1105.0020}.

\bibitem[{\citenamefont{Campbell et~al.}(2015)\citenamefont{Campbell, Ellis,
  and Giele}}]{Campbell:2015qma}
\bibinfo{author}{\bibfnamefont{J.~M.} \bibnamefont{Campbell}},
  \bibinfo{author}{\bibfnamefont{R.~K.} \bibnamefont{Ellis}}, \bibnamefont{and}
  \bibinfo{author}{\bibfnamefont{W.~T.} \bibnamefont{Giele}},
  \bibinfo{journal}{Eur. Phys. J.} \textbf{\bibinfo{volume}{C75}},
  \bibinfo{pages}{246} (\bibinfo{year}{2015}), \eprint{1503.06182}.

\bibitem[{\citenamefont{Boughezal et~al.}(2017)\citenamefont{Boughezal,
  Campbell, Ellis, Focke, Giele, Liu, Petriello, and
  Williams}}]{Boughezal:2016wmq}
\bibinfo{author}{\bibfnamefont{R.}~\bibnamefont{Boughezal}},
  \bibinfo{author}{\bibfnamefont{J.~M.} \bibnamefont{Campbell}},
  \bibinfo{author}{\bibfnamefont{R.~K.} \bibnamefont{Ellis}},
  \bibinfo{author}{\bibfnamefont{C.}~\bibnamefont{Focke}},
  \bibinfo{author}{\bibfnamefont{W.}~\bibnamefont{Giele}},
  \bibinfo{author}{\bibfnamefont{X.}~\bibnamefont{Liu}},
  \bibinfo{author}{\bibfnamefont{F.}~\bibnamefont{Petriello}},
  \bibnamefont{and} \bibinfo{author}{\bibfnamefont{C.}~\bibnamefont{Williams}},
  \bibinfo{journal}{Eur. Phys. J.} \textbf{\bibinfo{volume}{C77}},
  \bibinfo{pages}{7} (\bibinfo{year}{2017}), \eprint{1605.08011}.

\bibitem[{\citenamefont{Campbell and Neumann}(2019)}]{Campbell:2019dru}
\bibinfo{author}{\bibfnamefont{J.}~\bibnamefont{Campbell}} \bibnamefont{and}
  \bibinfo{author}{\bibfnamefont{T.}~\bibnamefont{Neumann}},
  \bibinfo{journal}{JHEP} \textbf{\bibinfo{volume}{12}}, \bibinfo{pages}{034}
  (\bibinfo{year}{2019}), \eprint{1909.09117}.

\bibitem[{\citenamefont{Angelescu and Huang}(2020)}]{Angelescu:2020yzf}
\bibinfo{author}{\bibfnamefont{A.}~\bibnamefont{Angelescu}} \bibnamefont{and}
  \bibinfo{author}{\bibfnamefont{P.}~\bibnamefont{Huang}}
  (\bibinfo{year}{2020}), \eprint{2006.16532}.

\bibitem[{\citenamefont{Corbett et~al.}(2018)\citenamefont{Corbett, Dolan,
  Englert, and Nordstr\"om}}]{Corbett:2017ecn}
\bibinfo{author}{\bibfnamefont{T.}~\bibnamefont{Corbett}},
  \bibinfo{author}{\bibfnamefont{M.~J.} \bibnamefont{Dolan}},
  \bibinfo{author}{\bibfnamefont{C.}~\bibnamefont{Englert}}, \bibnamefont{and}
  \bibinfo{author}{\bibfnamefont{K.}~\bibnamefont{Nordstr\"om}},
  \bibinfo{journal}{Phys. Rev. D} \textbf{\bibinfo{volume}{97}},
  \bibinfo{pages}{115040} (\bibinfo{year}{2018}), \eprint{1710.07530}.

\bibitem[{\citenamefont{Das~Bakshi et~al.}(2019)\citenamefont{Das~Bakshi,
  Chakrabortty, and Patra}}]{Bakshi:2018ics}
\bibinfo{author}{\bibfnamefont{S.}~\bibnamefont{Das~Bakshi}},
  \bibinfo{author}{\bibfnamefont{J.}~\bibnamefont{Chakrabortty}},
  \bibnamefont{and} \bibinfo{author}{\bibfnamefont{S.~K.} \bibnamefont{Patra}},
  \bibinfo{journal}{Eur. Phys. J.} \textbf{\bibinfo{volume}{C79}},
  \bibinfo{pages}{21} (\bibinfo{year}{2019}).

\bibitem[{\citenamefont{Ellis et~al.}(2020)\citenamefont{Ellis, Quevillon,
  Vuong, You, and Zhang}}]{Ellis:2020ivx}
\bibinfo{author}{\bibfnamefont{S.~A.~R.} \bibnamefont{Ellis}},
  \bibinfo{author}{\bibfnamefont{J.}~\bibnamefont{Quevillon}},
  \bibinfo{author}{\bibfnamefont{P.~N.~H.} \bibnamefont{Vuong}},
  \bibinfo{author}{\bibfnamefont{T.}~\bibnamefont{You}}, \bibnamefont{and}
  \bibinfo{author}{\bibfnamefont{Z.}~\bibnamefont{Zhang}}
  (\bibinfo{year}{2020}), \eprint{2006.16260}.

\bibitem[{\citenamefont{B\'elusca-Ma\"\i{}to
  et~al.}(2018)\citenamefont{B\'elusca-Ma\"\i{}to, Falkowski, Fontes, Rom\~ao,
  and Silva}}]{Belusca-Maito:2017iob}
\bibinfo{author}{\bibfnamefont{H.}~\bibnamefont{B\'elusca-Ma\"\i{}to}},
  \bibinfo{author}{\bibfnamefont{A.}~\bibnamefont{Falkowski}},
  \bibinfo{author}{\bibfnamefont{D.}~\bibnamefont{Fontes}},
  \bibinfo{author}{\bibfnamefont{J.~C.} \bibnamefont{Rom\~ao}},
  \bibnamefont{and} \bibinfo{author}{\bibfnamefont{J.~a.~P.}
  \bibnamefont{Silva}}, \bibinfo{journal}{JHEP} \textbf{\bibinfo{volume}{04}},
  \bibinfo{pages}{002} (\bibinfo{year}{2018}), \eprint{1710.05563}.

\bibitem[{\citenamefont{Giudice et~al.}(2007)\citenamefont{Giudice, Grojean,
  Pomarol, and Rattazzi}}]{Giudice:2007fh}
\bibinfo{author}{\bibfnamefont{G.~F.} \bibnamefont{Giudice}},
  \bibinfo{author}{\bibfnamefont{C.}~\bibnamefont{Grojean}},
  \bibinfo{author}{\bibfnamefont{A.}~\bibnamefont{Pomarol}}, \bibnamefont{and}
  \bibinfo{author}{\bibfnamefont{R.}~\bibnamefont{Rattazzi}},
  \bibinfo{journal}{JHEP} \textbf{\bibinfo{volume}{06}}, \bibinfo{pages}{045}
  (\bibinfo{year}{2007}), \eprint{hep-ph/0703164}.

\bibitem[{\citenamefont{Contino et~al.}(2013)\citenamefont{Contino, Ghezzi,
  Grojean, Muhlleitner, and Spira}}]{Contino:2013kra}
\bibinfo{author}{\bibfnamefont{R.}~\bibnamefont{Contino}},
  \bibinfo{author}{\bibfnamefont{M.}~\bibnamefont{Ghezzi}},
  \bibinfo{author}{\bibfnamefont{C.}~\bibnamefont{Grojean}},
  \bibinfo{author}{\bibfnamefont{M.}~\bibnamefont{Muhlleitner}},
  \bibnamefont{and} \bibinfo{author}{\bibfnamefont{M.}~\bibnamefont{Spira}},
  \bibinfo{journal}{JHEP} \textbf{\bibinfo{volume}{07}}, \bibinfo{pages}{035}
  (\bibinfo{year}{2013}), \eprint{1303.3876}.

\bibitem[{\citenamefont{Brivio et~al.}(2017)\citenamefont{Brivio, Jiang, and
  Trott}}]{Brivio:2017btx}
\bibinfo{author}{\bibfnamefont{I.}~\bibnamefont{Brivio}},
  \bibinfo{author}{\bibfnamefont{Y.}~\bibnamefont{Jiang}}, \bibnamefont{and}
  \bibinfo{author}{\bibfnamefont{M.}~\bibnamefont{Trott}},
  \bibinfo{journal}{JHEP} \textbf{\bibinfo{volume}{12}}, \bibinfo{pages}{070}
  (\bibinfo{year}{2017}), \eprint{1709.06492}.

\bibitem[{\citenamefont{Dedes et~al.}(2017)\citenamefont{Dedes, Materkowska,
  Paraskevas, Rosiek, and Suxho}}]{Dedes:2017zog}
\bibinfo{author}{\bibfnamefont{A.}~\bibnamefont{Dedes}},
  \bibinfo{author}{\bibfnamefont{W.}~\bibnamefont{Materkowska}},
  \bibinfo{author}{\bibfnamefont{M.}~\bibnamefont{Paraskevas}},
  \bibinfo{author}{\bibfnamefont{J.}~\bibnamefont{Rosiek}}, \bibnamefont{and}
  \bibinfo{author}{\bibfnamefont{K.}~\bibnamefont{Suxho}},
  \bibinfo{journal}{JHEP} \textbf{\bibinfo{volume}{06}}, \bibinfo{pages}{143}
  (\bibinfo{year}{2017}), \eprint{1704.03888}.

\bibitem[{\citenamefont{Golden and Randall}(1991)}]{Golden:1990ig}
\bibinfo{author}{\bibfnamefont{M.}~\bibnamefont{Golden}} \bibnamefont{and}
  \bibinfo{author}{\bibfnamefont{L.}~\bibnamefont{Randall}},
  \bibinfo{journal}{Nucl. Phys.} \textbf{\bibinfo{volume}{B361}},
  \bibinfo{pages}{3} (\bibinfo{year}{1991}).

\bibitem[{\citenamefont{Holdom and Terning}(1990)}]{Holdom:1990tc}
\bibinfo{author}{\bibfnamefont{B.}~\bibnamefont{Holdom}} \bibnamefont{and}
  \bibinfo{author}{\bibfnamefont{J.}~\bibnamefont{Terning}},
  \bibinfo{journal}{Phys. Lett.} \textbf{\bibinfo{volume}{B247}},
  \bibinfo{pages}{88} (\bibinfo{year}{1990}).

\bibitem[{\citenamefont{Altarelli and Barbieri}(1991)}]{Altarelli:1990zd}
\bibinfo{author}{\bibfnamefont{G.}~\bibnamefont{Altarelli}} \bibnamefont{and}
  \bibinfo{author}{\bibfnamefont{R.}~\bibnamefont{Barbieri}},
  \bibinfo{journal}{Phys. Lett.} \textbf{\bibinfo{volume}{B253}},
  \bibinfo{pages}{161} (\bibinfo{year}{1991}).

\bibitem[{\citenamefont{Peskin and Takeuchi}(1990)}]{Peskin:1990zt}
\bibinfo{author}{\bibfnamefont{M.~E.} \bibnamefont{Peskin}} \bibnamefont{and}
  \bibinfo{author}{\bibfnamefont{T.}~\bibnamefont{Takeuchi}},
  \bibinfo{journal}{Phys. Rev. Lett.} \textbf{\bibinfo{volume}{65}},
  \bibinfo{pages}{964} (\bibinfo{year}{1990}).

\bibitem[{\citenamefont{Grinstein and Wise}(1991)}]{Grinstein:1991cd}
\bibinfo{author}{\bibfnamefont{B.}~\bibnamefont{Grinstein}} \bibnamefont{and}
  \bibinfo{author}{\bibfnamefont{M.~B.} \bibnamefont{Wise}},
  \bibinfo{journal}{Phys. Lett.} \textbf{\bibinfo{volume}{B265}},
  \bibinfo{pages}{326} (\bibinfo{year}{1991}).

\bibitem[{\citenamefont{Altarelli et~al.}(1992)\citenamefont{Altarelli,
  Barbieri, and Jadach}}]{Altarelli:1991fk}
\bibinfo{author}{\bibfnamefont{G.}~\bibnamefont{Altarelli}},
  \bibinfo{author}{\bibfnamefont{R.}~\bibnamefont{Barbieri}}, \bibnamefont{and}
  \bibinfo{author}{\bibfnamefont{S.}~\bibnamefont{Jadach}},
  \bibinfo{journal}{Nucl. Phys.} \textbf{\bibinfo{volume}{B369}},
  \bibinfo{pages}{3} (\bibinfo{year}{1992}), \bibinfo{note}{[Erratum: Nucl.
  Phys.B376,444(1992)]}.

\bibitem[{\citenamefont{Peskin and Takeuchi}(1992)}]{Peskin:1991sw}
\bibinfo{author}{\bibfnamefont{M.~E.} \bibnamefont{Peskin}} \bibnamefont{and}
  \bibinfo{author}{\bibfnamefont{T.}~\bibnamefont{Takeuchi}},
  \bibinfo{journal}{Phys. Rev.} \textbf{\bibinfo{volume}{D46}},
  \bibinfo{pages}{381} (\bibinfo{year}{1992}).

\bibitem[{\citenamefont{Burgess et~al.}(1994)\citenamefont{Burgess, Godfrey,
  Konig, London, and Maksymyk}}]{Burgess:1993mg}
\bibinfo{author}{\bibfnamefont{C.~P.} \bibnamefont{Burgess}},
  \bibinfo{author}{\bibfnamefont{S.}~\bibnamefont{Godfrey}},
  \bibinfo{author}{\bibfnamefont{H.}~\bibnamefont{Konig}},
  \bibinfo{author}{\bibfnamefont{D.}~\bibnamefont{London}}, \bibnamefont{and}
  \bibinfo{author}{\bibfnamefont{I.}~\bibnamefont{Maksymyk}},
  \bibinfo{journal}{Phys. Lett.} \textbf{\bibinfo{volume}{B326}},
  \bibinfo{pages}{276} (\bibinfo{year}{1994}), \eprint{hep-ph/9307337}.

\bibitem[{\citenamefont{Grojean et~al.}(2013)\citenamefont{Grojean, Jenkins,
  Manohar, and Trott}}]{Grojean:2013kd}
\bibinfo{author}{\bibfnamefont{C.}~\bibnamefont{Grojean}},
  \bibinfo{author}{\bibfnamefont{E.~E.} \bibnamefont{Jenkins}},
  \bibinfo{author}{\bibfnamefont{A.~V.} \bibnamefont{Manohar}},
  \bibnamefont{and} \bibinfo{author}{\bibfnamefont{M.}~\bibnamefont{Trott}},
  \bibinfo{journal}{JHEP} \textbf{\bibinfo{volume}{04}}, \bibinfo{pages}{016}
  (\bibinfo{year}{2013}), \eprint{1301.2588}.

\bibitem[{\citenamefont{Englert and Spannowsky}(2015)}]{Englert:2014cva}
\bibinfo{author}{\bibfnamefont{C.}~\bibnamefont{Englert}} \bibnamefont{and}
  \bibinfo{author}{\bibfnamefont{M.}~\bibnamefont{Spannowsky}},
  \bibinfo{journal}{Phys. Lett.} \textbf{\bibinfo{volume}{B740}},
  \bibinfo{pages}{8} (\bibinfo{year}{2015}), \eprint{1408.5147}.

\bibitem[{\citenamefont{Bernlochner et~al.}(2019)\citenamefont{Bernlochner,
  Englert, Hays, Lohwasser, Mildner, Pilkington, Price, and
  Spannowsky}}]{Bernlochner:2018opw}
\bibinfo{author}{\bibfnamefont{F.~U.} \bibnamefont{Bernlochner}},
  \bibinfo{author}{\bibfnamefont{C.}~\bibnamefont{Englert}},
  \bibinfo{author}{\bibfnamefont{C.}~\bibnamefont{Hays}},
  \bibinfo{author}{\bibfnamefont{K.}~\bibnamefont{Lohwasser}},
  \bibinfo{author}{\bibfnamefont{H.}~\bibnamefont{Mildner}},
  \bibinfo{author}{\bibfnamefont{A.}~\bibnamefont{Pilkington}},
  \bibinfo{author}{\bibfnamefont{D.~D.} \bibnamefont{Price}}, \bibnamefont{and}
  \bibinfo{author}{\bibfnamefont{M.}~\bibnamefont{Spannowsky}},
  \bibinfo{journal}{Phys. Lett.} \textbf{\bibinfo{volume}{B790}},
  \bibinfo{pages}{372} (\bibinfo{year}{2019}), \eprint{1808.06577}.

\bibitem[{\citenamefont{Englert et~al.}(2019)\citenamefont{Englert, Giudice,
  Greljo, and Mccullough}}]{Englert:2019zmt}
\bibinfo{author}{\bibfnamefont{C.}~\bibnamefont{Englert}},
  \bibinfo{author}{\bibfnamefont{G.~F.} \bibnamefont{Giudice}},
  \bibinfo{author}{\bibfnamefont{A.}~\bibnamefont{Greljo}}, \bibnamefont{and}
  \bibinfo{author}{\bibfnamefont{M.}~\bibnamefont{Mccullough}},
  \bibinfo{journal}{JHEP} \textbf{\bibinfo{volume}{09}}, \bibinfo{pages}{041}
  (\bibinfo{year}{2019}), \eprint{1903.07725}.

\bibitem[{\citenamefont{Sirunyan
  et~al.}(2020{\natexlab{b}})}]{Sirunyan:2019wxt}
\bibinfo{author}{\bibfnamefont{A.~M.} \bibnamefont{Sirunyan}}
  \bibnamefont{et~al.} (\bibinfo{collaboration}{CMS}), \bibinfo{journal}{Eur.
  Phys. J.} \textbf{\bibinfo{volume}{C80}}, \bibinfo{pages}{75}
  (\bibinfo{year}{2020}{\natexlab{b}}), \eprint{1908.06463}.

\bibitem[{\citenamefont{Dawson and Giardino}(2020)}]{Dawson_2020}
\bibinfo{author}{\bibfnamefont{S.}~\bibnamefont{Dawson}} \bibnamefont{and}
  \bibinfo{author}{\bibfnamefont{P.~P.} \bibnamefont{Giardino}},
  \bibinfo{journal}{Physical Review D} \textbf{\bibinfo{volume}{101}}
  (\bibinfo{year}{2020}).

\bibitem[{\citenamefont{Alonso et~al.}(2014)\citenamefont{Alonso, Jenkins,
  Manohar, and Trott}}]{Alonso:2013hga}
\bibinfo{author}{\bibfnamefont{R.}~\bibnamefont{Alonso}},
  \bibinfo{author}{\bibfnamefont{E.~E.} \bibnamefont{Jenkins}},
  \bibinfo{author}{\bibfnamefont{A.~V.} \bibnamefont{Manohar}},
  \bibnamefont{and} \bibinfo{author}{\bibfnamefont{M.}~\bibnamefont{Trott}},
  \bibinfo{journal}{JHEP} \textbf{\bibinfo{volume}{04}}, \bibinfo{pages}{159}
  (\bibinfo{year}{2014}).

\bibitem[{\citenamefont{Murphy}(2018)}]{Murphy_2018}
\bibinfo{author}{\bibfnamefont{C.~W.} \bibnamefont{Murphy}},
  \bibinfo{journal}{Physical Review D} \textbf{\bibinfo{volume}{97}}
  (\bibinfo{year}{2018}).

\bibitem[{\citenamefont{Patra}(2019)}]{sunando_patra_2019_3404311}
\bibinfo{author}{\bibfnamefont{S.}~\bibnamefont{Patra}},
  \emph{\bibinfo{title}{sunandopatra/optex-1.0.0: Wo documentation}}
  (\bibinfo{year}{2019}),
  \urlprefix\url{https://doi.org/10.5281/zenodo.3404311}.

\bibitem[{\citenamefont{Baak et~al.}(2014)\citenamefont{Baak, Cúth, Haller,
  Hoecker, Kogler, Mönig, Schott, and Stelzer}}]{Baak_2014}
\bibinfo{author}{\bibfnamefont{M.}~\bibnamefont{Baak}},
  \bibinfo{author}{\bibfnamefont{J.}~\bibnamefont{Cúth}},
  \bibinfo{author}{\bibfnamefont{J.}~\bibnamefont{Haller}},
  \bibinfo{author}{\bibfnamefont{A.}~\bibnamefont{Hoecker}},
  \bibinfo{author}{\bibfnamefont{R.}~\bibnamefont{Kogler}},
  \bibinfo{author}{\bibfnamefont{K.}~\bibnamefont{Mönig}},
  \bibinfo{author}{\bibfnamefont{M.}~\bibnamefont{Schott}}, \bibnamefont{and}
  \bibinfo{author}{\bibfnamefont{J.}~\bibnamefont{Stelzer}},
  \bibinfo{journal}{The European Physical Journal C}
  \textbf{\bibinfo{volume}{74}} (\bibinfo{year}{2014}).

\bibitem[{\citenamefont{Aad et~al.}(2016{\natexlab{a}})}]{Khachatryan:2016vau}
\bibinfo{author}{\bibfnamefont{G.}~\bibnamefont{Aad}} \bibnamefont{et~al.}
  (\bibinfo{collaboration}{ATLAS, CMS}), \bibinfo{journal}{JHEP}
  \textbf{\bibinfo{volume}{08}}, \bibinfo{pages}{045}
  (\bibinfo{year}{2016}{\natexlab{a}}), \eprint{1606.02266}.

\bibitem[{\citenamefont{Aad et~al.}(2016{\natexlab{b}})}]{Aad:2015gba}
\bibinfo{author}{\bibfnamefont{G.}~\bibnamefont{Aad}} \bibnamefont{et~al.}
  (\bibinfo{collaboration}{ATLAS}), \bibinfo{journal}{Eur. Phys. J.}
  \textbf{\bibinfo{volume}{C76}}, \bibinfo{pages}{6}
  (\bibinfo{year}{2016}{\natexlab{b}}), \eprint{1507.04548}.

\bibitem[{\citenamefont{Aad et~al.}(2020{\natexlab{c}})}]{Aad:2019mbh}
\bibinfo{author}{\bibfnamefont{G.}~\bibnamefont{Aad}} \bibnamefont{et~al.}
  (\bibinfo{collaboration}{ATLAS}), \bibinfo{journal}{Phys. Rev. D}
  \textbf{\bibinfo{volume}{101}}, \bibinfo{pages}{012002}
  (\bibinfo{year}{2020}{\natexlab{c}}), \eprint{1909.02845}.

\bibitem[{\citenamefont{Aad et~al.}(2020{\natexlab{d}})}]{Aad:2020mkp}
\bibinfo{author}{\bibfnamefont{G.}~\bibnamefont{Aad}} \bibnamefont{et~al.}
  (\bibinfo{collaboration}{ATLAS}) (\bibinfo{year}{2020}{\natexlab{d}}),
  \eprint{2004.03447}.

\bibitem[{\citenamefont{Aad et~al.}(2020{\natexlab{e}})}]{Aad:2020plj}
\bibinfo{author}{\bibfnamefont{G.}~\bibnamefont{Aad}} \bibnamefont{et~al.}
  (\bibinfo{collaboration}{ATLAS}) (\bibinfo{year}{2020}{\natexlab{e}}),
  \eprint{2005.05382}.

\bibitem[{\citenamefont{Aad et~al.}(2019{\natexlab{a}})}]{ATLAS:2019ain}
\bibinfo{author}{\bibfnamefont{G.}~\bibnamefont{Aad}} \bibnamefont{et~al.}
  (\bibinfo{collaboration}{ATLAS}) (\bibinfo{year}{2019}{\natexlab{a}}),
  \eprint{ATLAS-CONF-2019-028}.

\bibitem[{\citenamefont{Aad et~al.}(2020{\natexlab{f}})}]{ATLAS:2020udg}
\bibinfo{author}{\bibfnamefont{G.}~\bibnamefont{Aad}} \bibnamefont{et~al.}
  (\bibinfo{collaboration}{ATLAS}) (\bibinfo{year}{2020}{\natexlab{f}}),
  \eprint{ATLAS-CONF-2020-007}.

\bibitem[{\citenamefont{Aaboud et~al.}(2018{\natexlab{a}})}]{Aaboud:2017jvq}
\bibinfo{author}{\bibfnamefont{M.}~\bibnamefont{Aaboud}} \bibnamefont{et~al.}
  (\bibinfo{collaboration}{ATLAS}), \bibinfo{journal}{Phys. Rev.}
  \textbf{\bibinfo{volume}{D97}}, \bibinfo{pages}{072003}
  (\bibinfo{year}{2018}{\natexlab{a}}), \eprint{1712.08891}.

\bibitem[{\citenamefont{Aaboud et~al.}(2018{\natexlab{b}})}]{Aaboud:2018urx}
\bibinfo{author}{\bibfnamefont{M.}~\bibnamefont{Aaboud}} \bibnamefont{et~al.}
  (\bibinfo{collaboration}{ATLAS}), \bibinfo{journal}{Phys. Lett. B}
  \textbf{\bibinfo{volume}{784}}, \bibinfo{pages}{173}
  (\bibinfo{year}{2018}{\natexlab{b}}), \eprint{1806.00425}.

\bibitem[{\citenamefont{Aad et~al.}(2019{\natexlab{b}})}]{Aad:2019lpq}
\bibinfo{author}{\bibfnamefont{G.}~\bibnamefont{Aad}} \bibnamefont{et~al.}
  (\bibinfo{collaboration}{ATLAS}), \bibinfo{journal}{Phys. Lett. B}
  \textbf{\bibinfo{volume}{798}}, \bibinfo{pages}{134949}
  (\bibinfo{year}{2019}{\natexlab{b}}), \eprint{1903.10052}.

\bibitem[{\citenamefont{Sirunyan et~al.}(2019)}]{Sirunyan:2018koj}
\bibinfo{author}{\bibfnamefont{A.~M.} \bibnamefont{Sirunyan}}
  \bibnamefont{et~al.} (\bibinfo{collaboration}{CMS}), \bibinfo{journal}{Eur.
  Phys. J. C} \textbf{\bibinfo{volume}{79}}, \bibinfo{pages}{421}
  (\bibinfo{year}{2019}), \eprint{1809.10733}.

\bibitem[{\citenamefont{Sirunyan
  et~al.}(2020{\natexlab{c}})}]{Sirunyan:2019qia}
\bibinfo{author}{\bibfnamefont{A.~M.} \bibnamefont{Sirunyan}}
  \bibnamefont{et~al.} (\bibinfo{collaboration}{CMS}), \bibinfo{journal}{JHEP}
  \textbf{\bibinfo{volume}{03}}, \bibinfo{pages}{131}
  (\bibinfo{year}{2020}{\natexlab{c}}), \eprint{1912.01662}.

\bibitem[{\citenamefont{Kumar and Martin}(2015)}]{Kumar:2015tna}
\bibinfo{author}{\bibfnamefont{N.}~\bibnamefont{Kumar}} \bibnamefont{and}
  \bibinfo{author}{\bibfnamefont{S.~P.} \bibnamefont{Martin}},
  \bibinfo{journal}{Phys. Rev. D} \textbf{\bibinfo{volume}{92}},
  \bibinfo{pages}{115018} (\bibinfo{year}{2015}), \eprint{1510.03456}.

\bibitem[{\citenamefont{Huang et~al.}(2020)\citenamefont{Huang, Morais, and
  Santos}}]{Huang:2020zde}
\bibinfo{author}{\bibfnamefont{D.}~\bibnamefont{Huang}},
  \bibinfo{author}{\bibfnamefont{A.~P.} \bibnamefont{Morais}},
  \bibnamefont{and} \bibinfo{author}{\bibfnamefont{R.}~\bibnamefont{Santos}}
  (\bibinfo{year}{2020}), \eprint{2009.09228}.

\bibitem[{\citenamefont{Cirigliano et~al.}(2019)\citenamefont{Cirigliano,
  Crivellin, Dekens, de~Vries, Hoferichter, and
  Mereghetti}}]{Cirigliano:2019vfc}
\bibinfo{author}{\bibfnamefont{V.}~\bibnamefont{Cirigliano}},
  \bibinfo{author}{\bibfnamefont{A.}~\bibnamefont{Crivellin}},
  \bibinfo{author}{\bibfnamefont{W.}~\bibnamefont{Dekens}},
  \bibinfo{author}{\bibfnamefont{J.}~\bibnamefont{de~Vries}},
  \bibinfo{author}{\bibfnamefont{M.}~\bibnamefont{Hoferichter}},
  \bibnamefont{and}
  \bibinfo{author}{\bibfnamefont{E.}~\bibnamefont{Mereghetti}},
  \bibinfo{journal}{Phys. Rev. Lett.} \textbf{\bibinfo{volume}{123}},
  \bibinfo{pages}{051801} (\bibinfo{year}{2019}), \eprint{1903.03625}.

\end{thebibliography}

\end{document}